# Graphics Processing Unit Accelerated Lattice Boltzmann Method Simulations of Dilute Gravity Currents

Damilola Adekanye[a], Amirul Khan[b], Alan Burns[c], William McCaffrey[d], Martin Geier[e], Martin Schönherr[e], Robert Dorrell[f]

[a]EPSRC Centre for Doctoral Training in Fluid Dynamics, School of Computing, University of Leeds, Leeds, LS2 9JT, United Kingdom

[b]School of Civil Engineering, University of Leeds, Leeds, LS2 9JT, United Kingdom

[c]School of Chemical and Process Engineering, University of Leeds, Leeds, LS2 9JT, United Kingdom

[d]School of Earth and Environment, University of Leeds, Leeds, UK, LS2 9JT, United Kingdom

[e]Institute for Computational Modelling in Civil Engineering, TU Braunschweig, Pockelsstr. 3, 38106 Brunswick, Germany

[f]Energy and Environment Institute, University of Hull, Hull HU6 7RX, United Kingdom

Corresponding author: Damilola Adekanye; Email: scda@leeds.ac.uk

## Abstract

Lattice Boltzmann method models offer a novel framework for the simulation of high Reynolds number dilute gravity currents. The numerical algorithm is well suited to acceleration via implementation on massively parallel computer architectures. Here we present two lattice Boltzmann method models of lock-exchange dilute gravity currents, in which the largest turbulent length scales are directly resolved. The three-dimensional simulations are accelerated by exporting computations to a graphics processing unit and are validated against experiments and high-resolution simulations for Reynolds numbers up to 30,000. The lattice Boltzmann method models achieve equivalent accuracy to conventional large eddy simulation models in the prediction of key flow properties. A conservative analysis of computational performance relative to conventional methods indicates that the presented framework reduces simulation times by two orders of magnitude. Therefore, it can be used as a foundation for the development of depth-resolving models that capture more of the complexity of environmental gravity currents.

## Nomenclature

| | |
|---|---|
| $c$ | Velocity quantum, $c = \Delta x/\Delta t$ |
| $\boldsymbol{c}_{ijk}^{Qm}$ | Discrete set of $m$ velocities in a three-dimensional space where $i, j, k \in \{1, 0, -1\}$ |
| $C_{nrs}$ | Cumulant of order $n+r+s$ |
| $C_{nrs}^{*}$ | Post collision cumulant of order $n+r+s$ |
| $C_{nrs}^{eq}$ | Equilibrium cumulant of order $n+r+s$ |
| $c_s$ | Speed of sound in the lattice |
| $C_s$ | Smagorinsky constant |
| CPU | Central processing unit |
| $D$ | Diffusivity of the scalar field |
| $DnQm$ | Defines a lattice structure with $n$ dimensions and $m$ velocities |
| DNS | Direct numerical simulation |
| $e_{L_1}$ | $L_1$ error |
| $\boldsymbol{e}^g$ | Unit vector in the direction of gravitational acceleration i.e., $\boldsymbol{e}^g = (0,0,-1)^T$ |
| $f$ | Continuous particle distribution function |
| $\boldsymbol{F}_B$ | Boussinesq forcing term |
| $f_{ijk}$ | Discretized particle distribution function |
| $f_{ijk}^{eq}$ | Discretized equilibrium particle distribution function |
| $F_w$ | Total frictional force applied to the lower boundary |
| $F_w^{LBM-LES}$ | Total frictional force applied to the lower boundary in an LBM-LES simulation |
| $F_w^{DNS}$ | Total frictional force applied to the lower boundary in a DNS simulation |
| $Fr$ | Froude number |
| $\boldsymbol{G}$ | Macroscopic body force acting on the flow |
| $g$ | Gravitational acceleration |
| $g'$ | Reduced gravity |
| GPU | Graphics processing unit |
| $H$ | Channel depth |
| $h_0$ | Initial current depth |
| $h_f$ | Height of the current head |
| $L_1$ | Length of the computational domain in the x direction |
| $L_2$ | Length of the computational domain in the y direction |
| $L_3$ | Length of the computational domain in the z direction |
| LBM | Lattice Boltzmann method |
| LBM-GPU | Lattice Boltzmann method solver that ports computations to a graphics processing unit |
| LBM-LES | Lattice Boltzmann method model that uses a large eddy simulation turbulence model |
| LES | Large eddy simulation |
| $M_{nrs}$ | Factorized central moments of order $n+r+s$ |
| $M_{nrs}^{*}$ | Post collision factorized central moments of order $n+r+s$ |
| $m_{\mathrm{ABC}}$ | Raw moments of order $A+B+C$ |
| MNUPS | Million node updates per second |
| $N_{L_3}$ | Number of nodes discretizing the $L_3$ dimension of the computational domain |
| $N_{nodes}$ | Total number of nodes in the computational domain |
| $N_t$ | Number of timesteps for which a results file was written to the hard drive |
| NS | Navier-Stokes |
| NS-DNS | A method that solves the Navier-Stokes equations and resolves all scales of the turbulent flow |
| NS-LES | A method that solves the Navier-Stokes equations and uses a large eddy simulation turbulence model |
| $P$ | Pressure |
| $P_j$ | A list of $n$ processors i.e., $P_j \in \{P_1, P_2, \dots, P_n\}$. |
| $p_k$ | Kinematic pressure |
| $Re_b$ | Buoyancy Reynolds number |
| $Re_{cr}$ | Critical buoyancy Reynolds number |
| $\bar{S}$ | Local stress tensor |
| $S_{ijk}$ | Momentum density source term in the lattice Boltzmann equation |
| $Sc$ | Schmidt number |
| $Sc_T$ | Turbulent Schmidt number |
| $t$ | Time |
| $t_{P_j}^{i}$ | A block of processing time on one of $n$ processors $P_j \in \{P_1, P_2, \dots, P_n\}$ |
| $T_{CPU}$ | Total central processing unit time required to run a simulation |
| $T_E$ | Total elapsed time required to run a simulation |
| $T_{End}$ | Time at which the simulation is terminated |
| $t_i$ | A list of $N_t$ times at which a results file was output i.e., $t_i \in \{t_1, t_2, \dots, t_{N_t}\}$ |
| $t_{IV}$ | Time at which a lock-exchange gravity current transitions from the inertial to viscous phase |
| $t_{SI}$ | Time at which a lock-exchange gravity current transitions from the slumping to inertial phase |
| $t_{SV}$ | Time at which a lock-exchange gravity current transitions from the slumping to viscous phase |
| $\boldsymbol{u}$ | Macroscopic velocity field $\boldsymbol{u} = (u, v, w)^T$ |
| $U_b$ | Buoyancy velocity |
| $u_f$ | Front velocity of the gravity current |
| $u_\tau$ | Friction velocity |
| $w_{ijk}^{Qm}$ | A constant set of weights corresponding to $m$ discrete velocities |
| $\boldsymbol{x}$ | Position in cartesian coordinate system i.e., $\boldsymbol{x} = (x, y, z)^T$ |
| $x_0$ | The distance of the lock gate from the start of the channel |
| $x_f$ | The distance between the current front and the initial lock gate position |
| $z_{max}^+$ | The maximum $z^+$ recorded across all wall-adjacent nodes throughout the duration of the simulation |
| $z^+$ | Non-dimensional distance from the wall |
| $\alpha$ | Coefficient of expansion |
| $\Delta$ | Filter width |
| $\Delta t$ | Time step |
| $\Delta x$ | Grid spacing |
| $\Delta z_1$ | Distance from a wall adjacent node to the boundary wall |
| $\Delta z_2$ | Distance from a wall adjacent node to the nearest node in the wall normal direction |
| $\delta\rho$ | The component of fluid density that fluctuates around the mean |
| $\kappa_{nrs}$ | Central moments of order $n + r + s$ |
| $\nu$ | Viscosity of the ambient fluid |
| $\nu_T$ | Local eddy viscosity |
| $\boldsymbol{\xi}$ | Particle velocity field $\boldsymbol{\xi} = (\xi_x, \xi_y, \xi_z)$ |
| $\boldsymbol{\Xi}$ | Velocity-frequency variable $\boldsymbol{\Xi} = \{\Xi, Y, Z\}$ |
| $\xi_{IP}$ | Empirical constant that parameterizes the inertial phase scaling law |
| $\xi_{VP}$ | Empirical constant that parameterizes the viscous phase scaling law |
| $\rho$ | Density |
| $\bar{\bar{\rho}}$ | Mean density |
| $\rho_0$ | Initial current density |
| $\rho_a$ | Density of the ambient fluid |
| $\rho_s$ | Solute density |
| $\tau$ | Characteristic relaxation time of the fluid |
| $\tau_w$ | Wall shear stress |
| $\tau_\Phi$ | Characteristic relaxation time of the scalar field |
| $\Phi$ | Continuous particle distribution function for the scalar field |
| $\Phi_{ijk}$ | Discretized particle distribution function for the scalar field |
| $\Phi_{ijk}^{eq}$ | Discretized equilibrium particle distribution function for the scalar field |
| $\chi$ | Solute concentration |
| $\chi_0$ | Initial solute concentration in the current |
| $\chi_a$ | Solute concentration of the ambient fluid |
| $\Omega$ | Continuous collision operator |
| $\Omega_{ijk}$ | Discretized collision operator |
| $\omega_{nrs}$ | Characteristic relaxation frequency of order $n + r + s$ |
| $\hat{}$ accent: | Indicates that the variable has been non-dimensionalized by the length scale $h_0$ and the velocity scale $(g'h_0)^{\frac{1}{2}}$ |
| $\tilde{}$ accent: | Indicates that the variable has been non-dimensionalized by the length scale $H$ and the velocity scale $U_b = (g'H)^{\frac{1}{2}}$ |
| $\bar{}$ accent: | Indicates that the variable has been span-wise averaged i.e., along the $L_2$ direction |

## 1. Introduction

Gravity currents are flows driven by the buoyancy forces that arise due to the action of gravity on a density gradient within a fluid. The broad class of environmental buoyancy-driven flows includes thermohaline flows[1] and saline currents[2], which are driven by temperature and salt-concentration gradients, respectively. Temperature and salinity gradients can occur simultaneously within a system, resulting in double-diffusive gravity currents.[3] Density gradients may also be generated by the suspension of particles within the flow, as is the case in turbidity currents, which are ocean-floor underflows consisting of a dense mixture of fluid and sediment particles.[4] In the case of turbidity currents, there is the added complexity of the current exchanging particulate material with the boundary through erosion and deposition, resulting in a two-way coupled relationship between the hydrodynamics of the flow and morphodynamics of the channel geometry.[5,6] Direct observation and measurement of turbidity currents in the environment are rare due to the infrequent and destructive nature of the flow. Therefore, the dynamics are investigated through theoretical, experimental, and numerical modelling. The propagation of gravity currents is of broad interest in environmental fluid dynamics, with relevance to research areas as diverse as the study of ocean current dynamics[7], and the development of carbon capture and storage processes[8].

The lock-exchange saline gravity current experiment is the classical problem used to investigate the dynamics of dilute gravity currents. The conventional experimental set-up is illustrated in Figure 1, and consists of a straight channel of depth $H$ in which a relatively light ambient fluid of density $\rho_a$, is separated by a gate from a fluid of density $\rho_0 > \rho_a$, and depth $h_0$. In the case of dilute gravity currents the density difference is small, $(\rho_0 - \rho_a)/\rho_a \ll 1$, hence the Boussinesq approximation can be applied, which assumes density variations are small enough to be neglected in the governing equations unless they are acted on by gravity.[9] The reduced gravity of the system is defined as $g' = g\,(\rho_0 - \rho_a)/\rho_a$, where $g$ is acceleration due to gravity.

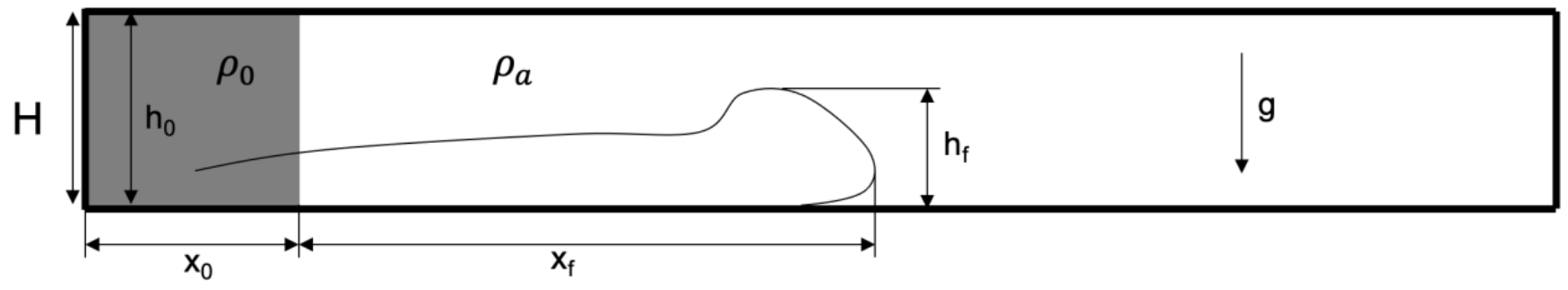

*Figure 1: Lock-exchange gravity current parameters*

The gate is placed at a distance $x_0$ from the start of the channel, and when removed, a gravity current is initiated by the resulting horizontal hydrostatic pressure gradient. The current propagates along the channel, with its front located at a distance $x_f(t)$ from the initial position of the gate ($x_0$), and at a height of $h_f(t)$ above the lower boundary, where t denotes time. The channel depth ($H$) is taken as the characteristic length scale of the flow, and the characteristic velocity scale is the buoyancy velocity $U_b = \sqrt{g'H}$. These characteristic scales combine with the viscosity of the ambient fluid ($\nu$) to define the buoyancy Reynolds number, $Re_b = U_b H/\nu$.

Following a brief period of frontal acceleration, lock-exchange gravity currents exhibit three distinct phases; the slumping, inertial and viscous phases.[10] In the slumping phase, the current front advances at a constant velocity ($u_f$) under a balance of pressure and drag forces.[10–13] The Froude number of the current is defined as the non-dimensional front velocity in the slumping phase, $Fr = \frac{u_f}{U_b}$.

In the inertial phase, the flow is governed by a balance of inertial and buoyancy forces. It has been determined, through both theoretical modelling and empirical study, that in the inertial phase front location ($x_f$) and velocity ($u_f$) tend asymptotically towards Equations 1 and 2 respectively.[10,14–16] In Equations 1-2, and throughout this paper, symbols with a ~ above them indicate values non-dimensionalized by the characteristic length scale $H$, and velocity scale $U_b$.

$$\tilde{x}_f = \xi_{IP}\left(\tilde{h}_0\tilde{x}_0\right)^{\frac{1}{3}}\tilde{t}^{\frac{2}{3}} \tag{1}$$

$$\tilde{u}_f = \frac{2}{3}\xi_{IP}\left(\tilde{h}_0\tilde{x}_0\right)^{\frac{1}{3}}\tilde{t}^{-\frac{1}{3}} \tag{2}$$

The viscous phase is governed by the balance of viscous and buoyancy forces. Hoult[15] derived a self-similarity solution of the depth-averaged Navier-Stokes equations to determine the viscous spreading rate of oil slicks on the surface of fresh water. Huppert[17] completed a similar analysis for the case of a dense viscous underflow over a no-slip boundary, which bears a closer relation to the lock-exchange problem, arriving at the scaling laws in Equations 3-4.

$$\tilde{x}_f = \xi_{VP}\left(\tilde{h}_0^3\,\tilde{x}_0^3 Re_b\right)^{\frac{1}{5}}\tilde{t}^{\frac{1}{5}} \tag{3}$$

$$\tilde{u}_f = \frac{1}{5}\xi_{VP}\left(\tilde{h}_0^3\,\tilde{x}_0^3 Re_b\right)^{\frac{1}{5}}\tilde{t}^{-\frac{4}{5}} \tag{4}$$

The empirical constant $\xi_{IP}$ was originally set to $\xi_{IP} = 1.6$ by Hoult[15], and $\xi_{IP} = 1.47$ by Huppert and Simpson[10]. Huppert[17] set the viscous phase constant to $\xi_{VP} = 1.13$. The values of the empirical constants were revised by Cantero et al.[18], who fit the $\xi_{IP}$ and $\xi_{VP}$ using a range of experimental and numerical results, ultimately arriving at the best fit values of $\xi_{IP} = 1.47$ and $\xi_{VP} = 3.2$.

The time of transition between the phases can be determined by equating the scaling laws[18]. Doing so results in the transition from the slumping to inertial phase occurring at $\tilde{t}_{SI}$, the transition from slumping to viscous occurring at $\tilde{t}_{SV}$, and the inertial to viscous phase transition occurring at $\tilde{t}_{IV}$, as shown in Equations 5-8 respectively. In the event that $\tilde{t}_{SV} < \tilde{t}_{SI}$ the flow will transition straight from the slumping to viscous phase and bypass the inertial phase entirely. The parameterization of the constants $\xi_{IP}$ and $\xi_{VP}$, conducted by Cantero et al.[18], had the effect of fitting the phase transition times predicted by the scaling laws to empirical data. The critical Reynolds number, above which the inertial phase develops, can be found by equating Equations 5 and 6, giving Equation 8.

$$\tilde{t}_{SI} = \left(\frac{2\xi_{IP}}{3Fr}\right)^3 \tilde{h}_0 \tilde{x}_0 \tag{5}$$

$$\tilde{t}_{SV} = \left(\frac{1}{5}\xi_{VP}\right)^{\frac{5}{4}} \frac{\left(\tilde{h}_0 \tilde{x}_0\right)^{\frac{3}{4}}}{Fr^{\frac{5}{4}}} Re_b^{\frac{1}{4}} \tag{6}$$

$$\tilde{t}_{IV} = \left(\frac{3\xi_{VP}}{10\xi_{IP}}\right)^{\frac{15}{7}} \left(\tilde{h}_0 \tilde{x}_0\right)^{\frac{4}{7}} Re_b^{\frac{3}{7}} \tag{7}$$

$$Re_{cr} = \left(\frac{2\xi_{IP}}{3Fr}\right)^{12} \left(\frac{5\mathrm{Fr}}{\xi_{VP}}\right)^5 \tilde{h}_0 \tilde{x}_0 \tag{8}$$

The time of transition from the slumping to inertial phase is a function of the initial non-dimensional lock length $\tilde{x}_0$. It is often convenient to alter the non-dimensionalization of velocity and length scales, such that $\hat{u} = \tilde{u}_f/(\tilde{h}_0)^{\frac{1}{2}} = u_f/(g'h_0)^{\frac{1}{2}}$ and $\hat{t} = \tilde{t}(\tilde{h}_0)^{\frac{1}{2}}/\tilde{x}_0 = t(g'h_0)^{\frac{1}{2}}/x_0$, which causes the transition time to collapse down to $\hat{t}_{SI} = \left(2\xi_{IP}\tilde{h}_0^{\frac{1}{2}}/3Fr\right)^3$.[18]

A range of depth-resolving models have been developed to simulate the mechanics of dilute gravity currents. The models numerically solve the incompressible Navier-Stokes (NS) equations of mass and momentum conservation coupled, via the Boussinesq approximation, with an advection-diffusion equation for the scalar concentration field.[19] As the dynamics of environmental gravity currents are substantially influenced by turbulent processes, such as turbulent mixing within the current and at the current-ambient interface, depth-resolving models are classified by the extent to which they resolve or model the turbulent length scales of the flow.

Numerical studies using direct numerical simulation (DNS), in which all turbulent length scales are resolved, have been conducted for lock-exchange gravity current flows with low to moderate Reynolds numbers.[18,20–26] However, due to the high levels of grid refinement required for DNS simulations, the approach is extremely computationally expensive, which precludes the study of highly turbulent flows in large complex domains of the sort observed in the environment.

Reynolds-averaged Navier-Stokes (RANS) models, derived by time-averaging of the governing equations, have been applied to the simulation of dilute gravity currents with both fixed and deformable boundaries.[27–32] Although RANS models are far less computationally expensive than DNS models, which has enabled the simulation of environmental scale flows, the inherent averaging of the governing equations reduces the accuracy with which they can capture the complex time-dependent turbulent flow features of gravity currents.

In large-eddy simulation (LES) models, the largest length scales are directly resolved, while the sub-grid scales of turbulent motion are modelled, usually using variants of the Smagorinsky model.[33] The grid resolution requirements for a well resolved gravity current simulation have been studied by Pelmard et al.[34]. NS-LES models benefit from enhanced accuracy relative to RANS approaches as they directly capture the large-scale turbulent flow features, whilst offering a less computationally expensive alternative to DNS.[34–39] However, the computational expense of conventional NS-LES models is still considerable, limiting the insights that can be gained into the dynamics of dilute gravity currents, as it becomes prohibitively expensive to extend the models to incorporate more of the full complexity of real-world environmental flows.

The lattice Boltzmann method (LBM) offers an alternative numerical framework to the aforementioned numerical methods, which all model fluid motion at the macroscopic scale, where the fluid and flow properties are continuous. The LB method differs fundamentally, as fluid motion is modelled at the mesoscopic scale, between macro and microscopic, where the fluid is described by a particle distribution function.[40] The evolution of the particle distribution function is governed by the Boltzmann equation, which is discretized to form an explicit numerical scheme. Guo, Shi and Zheng (2002) Guo et al.[41] first demonstrated that an LBM formulation equivalent to the Navier-Stokes equations, using the Boussinesq approximation to couple the governing equations, could effectively simulate flows driven by density gradients.

LBM models have been formulated to solve the depth-averaged shallow water equations, to predict the current height and front velocity of gravity currents. [42] However, there is very limited research on the application of LBM models to the simulation of gravity currents, where the LBM formulation is equivalent to the Navier-Stokes equations, and the Boussinesq approximation is applied. Ottolenghi et al.[43] published the first study of such a model, comparing the results of two and three-dimensional LBM large-eddy simulations (LBM-LES) of lock-exchange saline gravity currents against experimental data. Two-dimensional simulations were run for Reynolds numbers of $Re_b \in \{1000,\ 5000,\ 10000,\ 30000\}$, and three-dimensional simulations were run for $Re_b \in \{1000, 5000, 10000\}$. Ottolenghi et al.[43] made an initial comparison between the results of LBM simulations and experiments, demonstrating reasonable agreement with their experiments and theoretical results in the prediction of some key flow characteristics, such as the Froude number, entrainment of ambient fluid, and lobe-cleft development at the head. However, further validation is required, especially against DNS results, to determine whether the LBM framework can be used as an accurate numerical tool for the simulation of gravity currents.

The present study aims to address this open question and establish the LBM-LES method as an accurate alternative numerical framework for studying dilute gravity currents. Two LBM-LES codes, RAFSINE and VirtualFluids, are validated against a wide range of experimental data and high-resolution simulations. The key model performance tests, previously unaddressed in the literature, include the careful validation of LBM model predictions of flow phase transition against experimental data and high-resolution simulations, study of LBM-LES model accuracy in the near-wall region, and a comparison of the LBM-LES model accuracy to conventional NS-LES models. The accuracy and stability of an LBM model are influenced by the details of its formulation. The Ottolenghi et al.[43] LBM model uses a single relaxation time collision operator, which is the most widely used, but also the simplest of the available options. The LBM model implemented in VirtualFluids uses more advanced collision operators, creating potential for greater accuracy and stability in the simulations.[44]

Additionally, the computational performance advantages of the LBM framework are also yet to be established. In the Ottolenghi et al.[43] study, the LBM model is implemented in an in-house code that utilizes OpenMP parallelization on CPU cores, resulting in simulation times ranging from 1 to 5 days on a six-core desktop machine. The authors acknowledge that this does not reflect the potential computational efficiency of the approach. LBM models are particularly well suited to implementation on massively parallel machines, such as graphic processing units (GPUs) as the numerical scheme only contains calculations with locally defined variables. This allows LBM algorithms to effectively utilize the architecture of the GPU. When simulations are run on central processing units (CPUs) with multi-core processors, the domain is divided between the various cores, which perform calculations in parallel and communicate when necessary. LBM model implementations on a GPU offer much greater potential for parallel fluid simulations than the CPU, as each node in the domain can be assigned to a different thread and stepped forward in time in parallel, resulting in orders of magnitude reductions in simulation times.[45] In the present study, simulations in RAFSINE and VirtualFluids are accelerated by exporting computations to a GPU. The computational performance of the LBM-GPU implementations is compared to conventional methods in the field to quantify the relative performance gains that are realized through the novel application of this framework.

The paper is structured as follows: the methods used to develop the LBM-LES models and the NS-DNS validation simulations are detailed in Section 2; results and discussion regarding model accuracy and computational performance are presented in Section 3; finally, conclusions are delivered in Section 4.

## 2. Methods

### 2.1. Macroscopic Governing Equations

The macroscopic governing equations of a dilute saline gravity current flow are those of mass and momentum conservation for an incompressible flow coupled, via the Boussinesq approximation, with an advection-diffusion equation for the scalar concentration field. In the Boussinesq limit density ($\rho(\boldsymbol{x},t)$) is a linear function of saline concentration and is defined in Equation 9, where $\alpha = (\rho_s - \rho_a)/\rho_a$, $\rho_s$ is solute density, and $\chi(\boldsymbol{x},t)$ is solute concentration. Therefore, the macroscopic body force acting on the flow is $\boldsymbol{G}$, defined in Equation 10, where $\boldsymbol{e}^g = (0,0,-1)^T$ is the unit vector in the direction of gravitational acceleration. The constant term in $\boldsymbol{G}$ is absorbed into the pressure term of the momentum equation as hydrostatic pressure, $\boldsymbol{\nabla}P = \boldsymbol{\nabla}(p_k(\boldsymbol{x},t) - g\rho_a z)$, where $p_k$ is kinematic pressure. Therefore, the flow is driven by the Boussinesq forcing term $\boldsymbol{F}_B$, Equation 11.

$$\rho(\boldsymbol{x},t) = \rho_a(1 + \alpha\chi(\boldsymbol{x},t)) \tag{9}$$

$$\boldsymbol{G} = g\rho_a\big(1 + \alpha\chi(\boldsymbol{x},t)\big)\boldsymbol{e}^g \tag{10}$$

$$\boldsymbol{F}_B = g\rho_a\alpha\chi(\boldsymbol{x},t)\boldsymbol{e}^g \tag{11}$$

Solute concentration in the ambient and dense fluids are set to zero ($\chi_a = 0$) and unity ($\chi_0 = 1$) respectively. The governing equations are non-dimensionalized using the characteristic length scale $H$, and velocity scale $U_b$, resulting in the non-dimensional incompressible mass and momentum conservation equations (Equations 12-13), and an advection-diffusion equation for the scalar concentration field $\tilde{\chi}(\boldsymbol{x},t) = \chi(\boldsymbol{x},t)/\chi_0$ (Equation 14). Non-dimensional density is defined by the equation $\tilde{\rho}(\boldsymbol{x},t) = (\rho(\boldsymbol{x},t) - \rho_a)/(\rho_0 - \rho_a)$.

$$\boldsymbol{\nabla} \cdot \tilde{\boldsymbol{u}} = 0 \tag{12}$$

$$\frac{\partial \tilde{\boldsymbol{u}}}{\partial \tilde{t}} + \tilde{\boldsymbol{u}} \cdot \tilde{\boldsymbol{\nabla}} \tilde{\boldsymbol{u}} = -\tilde{\boldsymbol{\nabla}} \tilde{P} + \frac{1}{Re_b} \tilde{\nabla}^2 \tilde{\boldsymbol{u}} + \tilde{\chi} \boldsymbol{e}^g \tag{13}$$

$$\frac{\partial \tilde{\chi}}{\partial \tilde{t}} + \tilde{\boldsymbol{u}} \cdot \tilde{\boldsymbol{\nabla}} \tilde{\chi} = \frac{1}{Re_b Sc} \nabla^2 \tilde{\chi} \tag{14}$$

The governing equations contain two non-dimensional numbers, the buoyancy Reynolds number ($Re_b = U_b H/\nu$), and the Schmidt number ($Sc = \nu/D$), which is the ratio between the viscosity of the ambient fluid and the diffusivity of the scalar concentration field.

### 2.2. The Lattice Boltzmann Method Framework

Conventional depth-resolving models directly discretize the macroscopic governing equations and solve them numerically. The lattice Boltzmann method models fluid motion at the mesoscopic scale, i.e. between the micro and macroscopic scales. An overview of LBM theory is provided below, but readers are referred to Kruger et al.[44] for a rigorous derivation.

At the mesoscopic scale, distribution functions are the key variable and are used to represent the properties of a group of particles. The particle distribution function is an extension of volumetric mass density to include density in particle velocity space $(\xi_x, \xi_y, \xi_z)$, hence in three-dimensions $f(\boldsymbol{x}, \boldsymbol{\xi}, t)$ has the units presented in Equation 15.

$$[f(\boldsymbol{x},\boldsymbol{\xi},t)] = kg \cdot \frac{1}{m^3} \cdot \frac{1}{\left(\frac{m}{s}\right)^3} = \frac{kgs^3}{m^6}$$

15

The particle distribution function is a function of the particle position vector $(\boldsymbol{x})$, particle velocity $(\boldsymbol{\xi})$, and time $(t)$. Therefore, the function returns the particle density within a specified velocity range at a given location and time. The total derivative of the particle distribution function produces the Boltzmann equation (Equation 16), which includes a source term $\Omega(f)$ to account for the collision and subsequent redistribution of particles. In the present study, the external body force term in the Boltzmann equation is the Boussinesq forcing term $\boldsymbol{F}_B$.

$$Df(\boldsymbol{x},\boldsymbol{\xi},t) = \partial_t f(\boldsymbol{x},\boldsymbol{\xi},t) + \boldsymbol{\xi} \cdot \boldsymbol{\nabla} f(\boldsymbol{x},\boldsymbol{\xi},t) + \boldsymbol{F}_B \cdot \partial_\xi f(\boldsymbol{x},\boldsymbol{\xi},t) = \Omega(f)$$

16

The Boltzmann equation with forces is discretized to produce the lattice Boltzmann equation (LBE) with a momentum density source term $(S_{ijk})$ in Equation 17. This is an expression for the unknown distribution function $f_{ijk}(\boldsymbol{x},t)$, which is defined at the nodes of a lattice structure. The lattice is defined using the naming structure $DnQm$, where $n$ indicates the number of spatial dimensions, and $m$ is the number of discrete velocities. Three lattice structures are shown in Figure 2, $D3Q6$, $D3Q19$, and $D3Q27$. Lattice nodes are connected by a set of discrete velocities $\left(\boldsymbol{c}_{ijk}^{Qm}\right)$, where the indices $i, j, k$ can take the values $\in \{1, 0, -1\}$, corresponding to each component of velocity in a Cartesian coordinate system. The lattice spacing is defined as $\Delta x$, and particles move between site locations in time $\Delta t$. The LBE is second order accurate in both space and time. [44]

$$f_{ijk}\left(\boldsymbol{x} + \boldsymbol{c}_{ijk}^{Qm}\Delta t, t + \Delta t\right) = f_{ijk}(\boldsymbol{x},t) + \Omega_{ijk}\left(f_{ijk}\right) + S_{ijk}$$

17

The governing equations for macroscopic fluid flow, mass and momentum conservation can be recovered from the Boltzmann equation using asymptotic analysis[46] or Taylor expansion[47]. The raw moments of the distribution function $(f_{ijk})$ are calculated using Equation 18, where the order of the moment is determined by the sum of its indices $A + B + C$. Macroscopic density and momentum density are calculated from the zeroth and first order raw moments of $f_{ijk}$ respectively, as illustrated in Equations 19-20.

$$m_{\mathrm{ABC}} = \sum_{i,j,k} i^{\mathrm{A}} j^{B} k^{C} f_{ijk} \tag{18}$$

$$\rho = m_{000} = \sum_{i,j,k} f_{ijk} \tag{19}$$

$$\rho\boldsymbol{u} = \rho(m_{100}, m_{010}, m_{001}) = \left( \sum_{i,j,k} i f_{ijk} + \frac{\Delta t}{2} F_B^x, \sum_{i,j,k} j f_{ijk} + \frac{\Delta t}{2} F_B^y, \sum_{i,j,k} k f_{ijk} + \frac{\Delta t}{2} F_B^z \right) \tag{20}$$

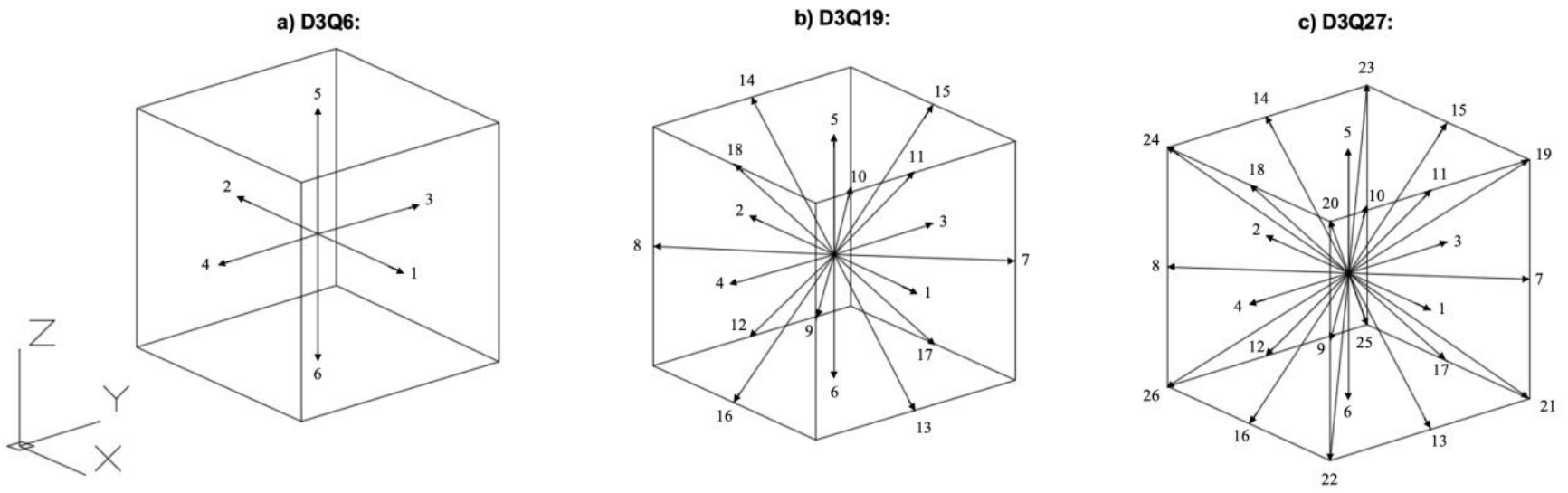

*Figure 2: Structure of a) D3Q6, b) D3Q19, and c) D3Q27 lattices.*

### 2.3. LBM Implementations

An LBM gravity current model has been implemented in two codes, RAFSINE and VirtualFluids, which accelerate simulations by exporting computations to a GPU device. The broad structure of the LBM-GPU implementations is illustrated through a flowchart in Figure 3. The flowchart emphasizes the transfer of data between the CPU and GPU and is broad enough to apply to both RAFSINE and VirtualFluids. Both packages are structured such that the pre-processing of the simulation i.e., the definition of simulation parameters and geometry is performed on the CPU, and then data is exported to the GPU where the computations are accelerated. The two models utilize different collision operators $(\Omega_{ijk})$, LES turbulence models, and GPU implementations, allowing for comparison of trade-offs in both accuracy and computational efficiency between the codes.

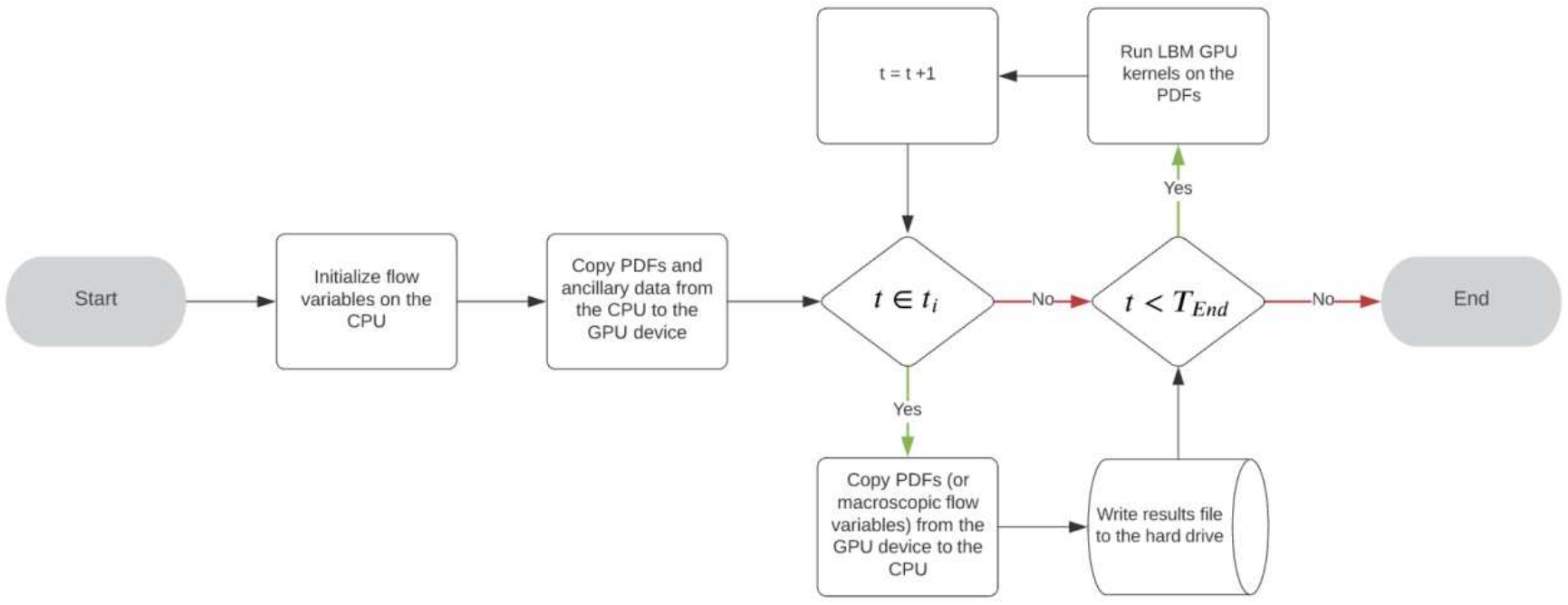

*Figure 3: Flowchart to outline the general structure of the LBM-GPU implementations in RAFSINE and VirtualFluids, with particular emphasis on the transfer of data between the CPU and GPU device.*

### 2.3.1. RAFSINE

RAFSINE was first developed by Delbosc[45] for the application of indoor airflow simulation. The version of the code applied in this study uses the BGK collision operator, which is also referred to as the single relaxation rate collision operator. The BGK approximation developed by Bhatnagar et al.[48] is widely used and is defined in Equation 21, where $f_{ijk}^{eq}$ is the equilibrium particle distribution function, Equation 22. Bhatnagar et al.[48] model the collision operator as the relaxation of the particle distribution function towards a state of local thermodynamic equilibrium.

$$\Omega_{ijk}(f) = \frac{1}{\tau}\left(f_{ijk}^{eq} - f_{ijk}\right) \tag{21}$$

The equilibrium is parametrized by the local velocity, density, and the speed of sound (related to the temperature), which is assumed to be constant in this lattice Boltzmann model. The Maxwellian equilibrium is simplified to a second order in velocity Taylor expansion, as shown in Equation 22.

$$f_{ijk}^{eq}(\boldsymbol{x}, t) = w_{ijk}^{Q19}\rho\left(1 + \frac{\boldsymbol{u}\cdot\boldsymbol{c}_{ijk}^{Q19}}{c_s^2} + \frac{\left(\boldsymbol{u}\cdot\boldsymbol{c}_{ijk}^{Q19}\right)^2}{2c_s^4} - \frac{\boldsymbol{u}\cdot\boldsymbol{u}}{2c_s^2}\right) \tag{22}$$

The velocity set $\boldsymbol{c}_{ijk}^{Q19}$ and the constant set of weights $w_{ijk}^{Q19}$ are defined in Equations 23-24 and correspond to the D3Q19 lattice, which is used to solve for the $f_{ijk}$ distribution in RAFSINE. The D3Q19 lattice is shown in Figure 2b.

$$\boldsymbol{c}_{ijk\in\{-1,0,1\}}^{Q19} = \begin{pmatrix} i \\ j \\ k \end{pmatrix} = \left[\begin{pmatrix} 0 \\ 0 \\ 0 \end{pmatrix}, \begin{pmatrix} 1 \\ 0 \\ 0 \end{pmatrix}, \begin{pmatrix} -1 \\ 0 \\ 0 \end{pmatrix}, \begin{pmatrix} 0 \\ 1 \\ 0 \end{pmatrix}, \begin{pmatrix} 0 \\ -1 \\ 0 \end{pmatrix}, \begin{pmatrix} 0 \\ 0 \\ 1 \end{pmatrix}, \begin{pmatrix} 0 \\ 0 \\ -1 \end{pmatrix}, \begin{pmatrix} 1 \\ 1 \\ 0 \end{pmatrix}, \begin{pmatrix} -1 \\ -1 \\ 0 \end{pmatrix}, \begin{pmatrix} 1 \\ -1 \\ 0 \end{pmatrix}, \begin{pmatrix} -1 \\ 1 \\ 0 \end{pmatrix}, \begin{pmatrix} 1 \\ 0 \\ 1 \end{pmatrix}, \begin{pmatrix} -1 \\ 0 \\ -1 \end{pmatrix}, \begin{pmatrix} 1 \\ 0 \\ -1 \end{pmatrix}, \begin{pmatrix} -1 \\ 0 \\ 1 \end{pmatrix}, \begin{pmatrix} 0 \\ 1 \\ 1 \end{pmatrix}, \begin{pmatrix} 0 \\ -1 \\ -1 \end{pmatrix}, \begin{pmatrix} 0 \\ 1 \\ -1 \end{pmatrix}, \begin{pmatrix} 0 \\ -1 \\ 1 \end{pmatrix}\right] \tag{23}$$

$$w_{ijk}^{Q19} \begin{cases} \frac{1}{3}, & \|\boldsymbol{c}_{ijk}^{Q19}\| = 0 \\ \frac{1}{18}, & \|\boldsymbol{c}_{ijk}^{Q19}\| = 1 \\ \frac{1}{36}, & \|\boldsymbol{c}_{ijk}^{Q19}\| = \sqrt{2} \end{cases} \tag{24}$$

The macroscopic viscosity $(\nu)$ is derived from the characteristic relaxation time of the fluid $(\tau)$, as shown in Equation 25, where $c_s = \sqrt{\Delta x^2/(3\Delta t^2)}$ is the speed of sound in the lattice.

$$\nu = c_s^2\left(\tau - \frac{1}{2}\right)\frac{\Delta x^2}{\Delta t} \tag{25}$$

The momentum density source term $S_{ijk}$ is defined in Equation 26.[44]

$$S_{ijk} = \left(1 - \frac{\Delta t}{2\tau}\right) w_{ijk}^{Q19}\left(\frac{\boldsymbol{c}_{ijk}^{Q19}}{c_s^2} + \frac{\left(\boldsymbol{c}_{ijk}^{Q19}\left(\boldsymbol{c}_{ijk}^{Q19}\right)^T - c_s^2\boldsymbol{\delta}\right)\boldsymbol{u}}{c_s^4}\right) \cdot \boldsymbol{F}_B \tag{26}$$

RAFSINE is capable of simulating density driven flows by coupling two LBM equations, one for the conservation of mass and momentum in the fluid, and a second for the advection and diffusion of the concentration field $(\chi)$. The particle distribution function for the concentration field is $\Phi_{ijk}$, and its zeroth order raw moment is the macroscopic concentration $\chi$, as shown in Equation 27. The equations are coupled via the Boussinesq approximation, as $\chi$ is used to update the local value of the forcing term $\boldsymbol{F}_B$ at a given time step.

$$\chi = \sum_{i,j,k} \Phi_{ijk} \tag{27}$$

The LBM-BGK equation for the $\Phi_{ijk}$ distribution is presented in Equation 28, where $\Phi_{ijk}^{eq}(\boldsymbol{x},t)$ is the equilibrium distribution, defined in Equation 29. The macroscopic diffusivity is determined by the relaxation time for the scalar field $\tau_\Phi$, as shown in Equation 30.

$$\Phi_{ijk}\left(\boldsymbol{x} + \boldsymbol{c}_{ijk}^{Q6}\Delta t, t + \Delta t\right) = \Phi_{ijk}(\boldsymbol{x},t) - \frac{1}{\tau_\Phi}\left(\Phi_{ijk}(\boldsymbol{x},t) - \Phi_{ijk}^{eq}(\boldsymbol{x},t)\right) \tag{28}$$

$$\Phi_{ijk}^{eq}(\boldsymbol{x},t) = w_{ijk}^{Q6}\chi\left(1 + \frac{\boldsymbol{u}\cdot\boldsymbol{c}_{ijk}^{Q6}}{c_s^2}\right) \tag{29}$$

$$D = c_s^2\left(\tau_\phi - \frac{1}{2}\right)\frac{\Delta x^2}{\Delta t} \tag{30}$$

A D3Q6 lattice (Figure 2a) is used to solve for the $\Phi_{ijk}$ distribution in RAFSINE, in line with the formulation originally used by Delbosc et al.[49]. The lattice has the velocity set $\boldsymbol{c}_{ijk}^{Q6}$, defined in Equation 31, and a constant weight $w_{ijk}^{Q6} = 1/6$.

$$\boldsymbol{c}_{ijk\in\{-1,0,1\}}^{Q6} = \begin{pmatrix} i \\ j \\ k \end{pmatrix} = \left[\begin{pmatrix} 1 \\ 0 \\ 0 \end{pmatrix}, \begin{pmatrix} -1 \\ 0 \\ 0 \end{pmatrix}, \begin{pmatrix} 0 \\ 1 \\ 0 \end{pmatrix}, \begin{pmatrix} 0 \\ -1 \\ 0 \end{pmatrix}, \begin{pmatrix} 0 \\ 0 \\ 1 \end{pmatrix}, \begin{pmatrix} 0 \\ 0 \\ -1 \end{pmatrix}\right] \tag{31}$$

The code utilizes a standard Smagorinsky turbulence model, which is an LES approach that models energy damping due to sub-grid turbulence through a local eddy viscosity $(\nu_T)$, such that $\nu = \nu_0 + \nu_T$, as given by Hou et al.[50].

$$\nu_T = C_s \Delta^2 |\bar{S}| \tag{32}$$

$$\bar{\boldsymbol{S}} = \frac{1}{2}(\boldsymbol{\nabla u} + \boldsymbol{\nabla}\mathbf{u}^{\mathrm{T}}) = \sum_{i,j,k} \boldsymbol{c}_{ijk}^{Q19} \cdot \boldsymbol{c}_{ijk}^{Q19} \left(f_{ijk} - f_{ijk}^{eq}\right) \tag{33}$$

$$|\bar{S}| = \frac{1}{6C_s\Delta^2}\left(\sqrt{\nu^2 + 18C_s^2\Delta^{2\sqrt{\bar{S}\bar{S}}}} - \nu\right) \tag{34}$$

The method assumes that small scale turbulence is isotropic and is implemented using Equation 32, where $C_s$ is the Smagorinsky constant, $\Delta$ is the filter width, and $\bar{S}$ is the local stress tensor, defined in Equation 33. The magnitude of the local stress tensor is calculated from Equation 34. The Smagorinsky constant is set to $C_s = 0.03$, according to the recommendations of Delbosc[45].

The diffusivity of the scalar field is also influenced by the effects of sub-grid turbulence. In the Smagorinsky model diffusivity is calculated using Equation 35, as determined by Liu et al.[51], where $Sc_T$ is the turbulent Schmidt number, which is taken to be equal to $C_s$.

$$D = c_s^2\left(\left(\tau_\Phi + \frac{C_s\Delta^2|\bar{S}|}{Sc_T}\right) - \frac{1}{2}\right)\frac{\Delta x^2}{\Delta t} \tag{35}$$

RAFSINE's LBM implementation contains several measures to maximize computational efficiency when running on GPUs. These include a number of adaptations to optimize the utilization of the GPU's memory bandwidth, such as eliminating redundant memory accesses, increasing data coalescence, and efficient reading/writing of distributions. Delbosc et al.[49] demonstrated that the optimizations resulted in a computational performance just 6% below the maximum capacity of the available hardware. When validated against experimental data of thermal flow in a $32\ m^3$ room, RAFSINE simulated the flow at 1.5 times real time on an NVIDIA Tesla C2070 GPU.

### 2.3.1. VirtualFluids

VirtualFluids was developed by the Institute for Computational Modelling in Civil Engineering (iRMB) at TU Braunschweig. VirtualFluids utilizes more advanced collision operators, namely the factorized central moments method to solve for the advection-diffusion of the scalar field, and the cumulant collision operator to solve for the conservation of mass and momentum. A D3Q27 lattice, shown in Figure 2c, is used to solve for both the $f_{ijk}$ and $\Phi_{ijk}$ distributions. The velocity set for the D3Q27 lattice is shown in Equation 36.

$$\boldsymbol{c}_{ijk\in\{-1,0,1\}}^{Q27} = \begin{pmatrix} i \\ j \\ k \end{pmatrix} = \left[\begin{pmatrix} 0 \\ 0 \\ 0 \end{pmatrix}, \begin{pmatrix} 1 \\ 0 \\ 0 \end{pmatrix}, \begin{pmatrix} -1 \\ 0 \\ 0 \end{pmatrix}, \begin{pmatrix} 0 \\ 1 \\ 0 \end{pmatrix}, \begin{pmatrix} 0 \\ -1 \\ 0 \end{pmatrix}, \begin{pmatrix} 0 \\ 0 \\ 1 \end{pmatrix}, \begin{pmatrix} 0 \\ 0 \\ -1 \end{pmatrix}, \begin{pmatrix} 1 \\ 1 \\ 0 \end{pmatrix}, \begin{pmatrix} -1 \\ -1 \\ 0 \end{pmatrix}, \begin{pmatrix} 1 \\ -1 \\ 0 \end{pmatrix},\right.$$
$$\begin{pmatrix} -1 \\ 1 \\ 0 \end{pmatrix}, \begin{pmatrix} 1 \\ 0 \\ 1 \end{pmatrix}, \begin{pmatrix} -1 \\ 0 \\ -1 \end{pmatrix}, \begin{pmatrix} 1 \\ 0 \\ -1 \end{pmatrix}, \begin{pmatrix} -1 \\ 0 \\ 1 \end{pmatrix}, \begin{pmatrix} 0 \\ 1 \\ 1 \end{pmatrix}, \begin{pmatrix} 0 \\ -1 \\ -1 \end{pmatrix}, \begin{pmatrix} 0 \\ 1 \\ -1 \end{pmatrix}, \begin{pmatrix} 0 \\ -1 \\ 1 \end{pmatrix},$$
$$\left.\begin{pmatrix} 1 \\ 1 \\ 1 \end{pmatrix}, \begin{pmatrix} 1 \\ -1 \\ 1 \end{pmatrix}, \begin{pmatrix} 1 \\ 1 \\ -1 \end{pmatrix}, \begin{pmatrix} 1 \\ -1 \\ -1 \end{pmatrix}, \begin{pmatrix} -1 \\ 1 \\ 1 \end{pmatrix}, \begin{pmatrix} -1 \\ -1 \\ 1 \end{pmatrix}, \begin{pmatrix} -1 \\ 1 \\ -1 \end{pmatrix}, \begin{pmatrix} -1 \\ -1 \\ -1 \end{pmatrix}\right] \tag{36}$$

The factorized central moments (FCM) method was applied to the advection-diffusion equation by Yang et al.[52]. Whilst the BGK collision operator performs collisions in momentum space, relaxing distributions towards their equilibrium state, the factorized central moments method performs collisions in factorized central moments space. Moments of the discrete distribution function $\Phi_{ijk}$ can be obtained by first converting $\Phi_{ijk}$ into a continuous function using the Dirac delta function $\delta$, as shown in Equation 37.

$$\Phi(\boldsymbol{\xi}) = \Phi(\xi_x + \xi_y + \xi_z) = \sum_{i,j,k} \Phi_{ijk}\delta(ic - \xi_x)\delta(jc - \xi_y)\delta(kc - \xi_z) \tag{37}$$

The central moment generating function is then obtained by applying a bi-lateral Laplace transform to the function $\Phi(\boldsymbol{u} - \boldsymbol{\xi})$, as shown in Equation 38. These are referred to as central moments, as $\Phi(\boldsymbol{\xi})$ has been shifted into the frame of reference of the macroscopic fluid moving with velocity $\boldsymbol{u}$.

$$F(\Xi) \coloneqq \mathcal{L}\{\Phi(\boldsymbol{u} - \boldsymbol{\xi})\}(\Xi)$$

$$= e^{-\Xi\cdot\boldsymbol{u}} \int_{-\infty}^{\infty} \Phi(\boldsymbol{\xi}) e^{\Xi\cdot\boldsymbol{\xi}} \boldsymbol{d\xi}$$

$$= e^{-\Xi\cdot\boldsymbol{u}} \sum_{i,j,k} \Phi_{ijk} \int_{-\infty}^{\infty} \delta(ic - \xi_x)\delta(jc - \xi_y)\delta(kc - \xi_z) e^{\Xi\cdot\boldsymbol{\xi}} \boldsymbol{d\xi}$$

$$= \sum_{i,j,k} \Phi_{ijk} e^{\Xi(\mathrm{ic}-\mathrm{u}) + Y(jc-v) + Z(kc-w)}$$

38

The moment generating function $F(\Xi)$, is a function of the velocity-frequency variable $\boldsymbol{\Xi} = \{\Xi, Y, Z\}$, and derivatives of $F(\Xi)$ produce non-orthogonal central moments of $\Phi_{ijk}$ of order $(n + r + s)$, as shown in Equation 39.

$$\kappa_{nrs} \coloneqq c^{-(n+r+s)} \frac{\partial^n \partial^r \partial^s}{\partial \Xi^n \partial Y^r \partial Z^s} F(\Xi)\Big|_{\Xi=0}$$

$$= \sum_{i,j,k} (i-u)^n (j-v)^r (k-w)^s \Phi_{ijk}$$

39

Statistical independence of the central moments can be achieved by factorizing them.[53] The factorized central moments are calculated using Equations 40-49, where brackets are used to denote permutations of the indices.[52]

$$M_{000} = \kappa_{000} = \chi$$

40

$$M_{(100)} = \kappa_{(100)}$$

41

$$M_{(110)} = \kappa_{(110)}$$

42

$$M_{111} = \kappa_{111}$$

43

$$M_{(200)} = \kappa_{(200)} - \frac{1}{3}\kappa_{000} \tag{44}$$

$$M_{(210)} = \kappa_{(210)} - \frac{1}{3}\kappa_{(010)} \tag{45}$$

$$M_{(220)} = \kappa_{(220)} - \frac{1}{3}\kappa_{000} \tag{46}$$

$$M_{(221)} = \kappa_{(221)} - \frac{1}{9}\kappa_{(001)} \tag{47}$$

$$M_{(211)} = \kappa_{(211)} - \frac{1}{3}\kappa_{(011)} \tag{48}$$

$$M_{222} = \kappa_{222} - \frac{1}{27}\kappa_{000} \tag{49}$$

Collision is then performed in moment space, where moments $M_{nrs}$ are relaxed at the frequency $\omega_{nrs}$, towards their equilibria, which for factorised central moments is set to zero. The post collision moments $M_{nrs}^*$ are calculated using Equation 50.

$$M_{nrs}^* = (1 - \omega_{nrs})M_{nrs} \tag{50}$$

When applied to the advection-diffusion equation, the only conserved moment is $M_{000} = \chi$. Diffusivity is calculated using the relaxation frequency of $\kappa_{(100)}$, as shown in Equation 51.

$$D = \frac{1}{3}\left(\frac{1}{\omega_{100}} - \frac{1}{2}\right)\frac{\Delta x^2}{\Delta t} \tag{51}$$

The VirtualFluids code solves the incompressible LBM equation in which fluid density is decomposed into its mean ($\bar{\bar{\rho}}$) and fluctuating ($\delta\rho$) components, as shown in Equation 52. The $\delta\rho$ component is calculated by the zeroth raw moment of the distribution $f_{ijk}$ (Equation 53), and the $\bar{\bar{\rho}} = 1$ in lattice units.

$$\rho = \bar{\bar{\rho}} + \delta\rho \tag{52}$$

$$\delta\rho = m_{000} = \sum_{i,j,k} f_{ijk} \tag{53}$$

The cumulant collision operator is used to solve for mass and momentum conservation in the fluid.[54] The cumulant operator performs collisions in cumulant space, where cumulants are statistically independent observable quantities of the momentum distribution $f_{ijk}$. They are calculated from the series expansion of the logarithm of the moment generating function $\mathcal{L}\{f(-\boldsymbol{\xi})\}(\boldsymbol{\Xi})$, as shown in Equation 54.

$$C_{nrs} \coloneqq c^{-(n+r+s)} \frac{\partial^n \partial^r \partial^s}{\partial \Xi^n \partial Y^r \partial Z^s} \ln\left(\mathcal{L}\{f(-\boldsymbol{\xi})\}(\boldsymbol{\Xi})\right)\Big|_{\Xi=0} \tag{54}$$

The post collision cumulants $C^*_{nrs}$ are calculated using Equation 55, where $C^{eq}_{nrs}$ is the equilibrium cumulant.

$$C^*_{nrs} = (1 - \omega_{nrs})C_{nrs} + \omega_{nrs}C^{eq}_{nrs} \tag{55}$$

The process for performing efficient transformations between momentum space and cumulant space is detailed by Geier et al.[55]. As the cumulant collision operator is applied to the conservation of mass and momentum, the zeroth and first order cumulants are conserved, which relate to density and velocity as shown in Equations 56-57.

$$\rho = C_{000} \tag{56}$$

$$\boldsymbol{u} = (u, v, w) = (C_{100}, C_{010}, C_{001}) \quad (57)$$

Viscosity is calculated using the relaxation rate of the second order cumulants, Equation 58.

$$\nu = \frac{1}{3}\left(\frac{1}{\omega_{110}} - \frac{1}{2}\right)\frac{\Delta x^2}{\Delta t} \quad (58)$$

A cumulant collision operator is deemed to be more accurate than the BGK collision operator due to the incorporation of higher order velocity terms in the equilibrium and its Galilean invariant viscosity.[54] The current study applies the parameterized cumulant method of Geier et al.[55]. In this the relaxation rates of the third order cumulants are chosen to eliminate the leading order error in diffusion such that the handling of viscosity becomes essentially fourth order accurate.

VirtualFluids does not incorporate a sub-grid eddy viscosity turbulence model which would destroy the advantage of the fourth order accuracy. To stabilize the method for resolutions not reaching DNS quality it is sufficient to add a limiter on the relaxation of the third order cumulants. Compared to adding an explicit sub-gird model the stabilized parametrized cumulant method has been shown to require half the resolution to obtain the same enstrophy production.[56] The method was also successful in accurately predicting the drag crisis behind a sphere[57], simulating flows with Reynolds numbers ranging from $200$ to $10^5$. The approach taken by the cumulant method towards turbulence is to provide the highest possible accuracy even at low resolution while non-resolved scales are naturally cut off. Adding a sub-grid model to the parameterized cumulant method has no known advantages and typically leads to inferior results.[56]

VirtualFluids has been optimized to run on GPUs, using indirect addressing to facilitate simulation in complex geometries, and the Esoteric Twist data structure to minimize memory overhead and traffic on the GPU.[58]

Previous studies in VirtualFluids have considered passive scalar transport in which the cumulant and FCM kernels are one-way coupled, so buoyancy forces do not drive the flow.[52,59] In the present study two-way coupling has been implemented to allow the simulation of a buoyancy driven gravity current flow.

### 2.4. Nek5000 Direct Numerical Simulations

Direct numerical simulations are run in the high-order solver Nek5000 to provide high-resolution simulation results against which to benchmark the LBM-LES codes, in addition to that already available in literature. Nek5000-v19.0 is a CFD solver developed by Argonne National Laboratory[60] based on the spectral-element method (SEM)[61]. The approach discretizes the computational domain into E elements, each containing an Nth-order polynomial discretization. In the present study 7th order polynomials were used for optimal accuracy and performance.[60] The non-dimensional governing equations outlined in Section 2.1 were solved in Nek5000 using 2nd order backward differential formula (BDF) and operator-integration factor scheme (OIFS) extrapolation, which allows for a target Courant number of 2-5 whilst maintaining stability and accuracy.[62] The residual tolerance for pressure was set to $10^{-9}$, while the tolerances for fluid velocity and concentration were set to $10^{-10}$.

### 2.5. Lock-Exchange Saline Gravity Current Model

The full list of test cases are presented in Table 1, where the channel dimensions $(L_1,\ L_2, L_3 = H)$ are defined in Figure 4. $N_{L_3}$ is the number of nodes used to resolve the channel depth $H$, and $N_{nodes}$ is the total number of mesh points in the computational domain.

The LBM-LES models are validated against the DNS results published by Cantero et al.[18,21], where simulations were run with buoyancy Reynolds numbers ranging from $Re_b \in [895, 15000]$. As DNS becomes prohibitively expensive at high Reynolds numbers, it was also necessary to validate against the experimental results of Ottolenghi et al.[43], where experiments were conducted for Reynolds numbers ranging from $Re_b \in [1000, 30000]$. To achieve a more robust validation, DNS simulations were run in Nek5000 for Cases 2 and 4 in Table 1.

| *Case No.* | $Re_b$ | $L_3/x_0$ | $L_1$ | $L_2$ | $N_{L_3}$ | $N_{nodes}$ $(10^6)$ | *Data Type* | *Data Source* |
|---|---|---|---|---|---|---|---|---|
| *1* | 895 | 1 | 25 | 1.5 | 100 | 37.5 | DNS | Cantero et al.[18] |
| *2* | 1000 | 1 | 15 | 1 | 100 | 15 | DNS; Exp | Nek5000; Ottolenghi et al. [43] |
| *3* | 3450 | 1 | 25 | 1.5 | 100 | 37.5 | DNS | Cantero et al.[18] |
| *4* | 5000 | 1 | 15 | 1 | 100 | 15 | DNS; Exp | Nek5000; Ottolenghi et al.[43] |
| *5* | 8950 | 1 | 25 | 1.5 | 100 | 37.5 | DNS | Cantero et al.[18] |
| *6* | 10000 | 1 | 15 | 1 | 100 | 15 | Exp | Ottolenghi et al.[43] |
| *7* | 15000 | 1 | 25 | 1.5 | 104 | 42.2 | DNS | Cantero et al.[21] |
| *8* | 30000 | 1 | 15 | 1 | 140 | 41.2 | Exp | Ottolenghi et al.[43] |

*Table 1: Saline current LBM-LES simulation parameters*

The geometry and boundary conditions used in Cases $\in \{2, 4, 6, 8\}$, outlined in Figure 4a, were selected to model the experimental conditions of Ottolenghi et al.[43]. A no-slip boundary condition for the velocity field and a no-flux boundary condition for the concentration field is applied on the upper and lower boundaries, end walls, and span-wise walls. The initial velocity was zero throughout the domain, while the concentration was set to $\tilde{\chi} = 1$ within the lock, and $\tilde{\chi} = 0$ elsewhere.

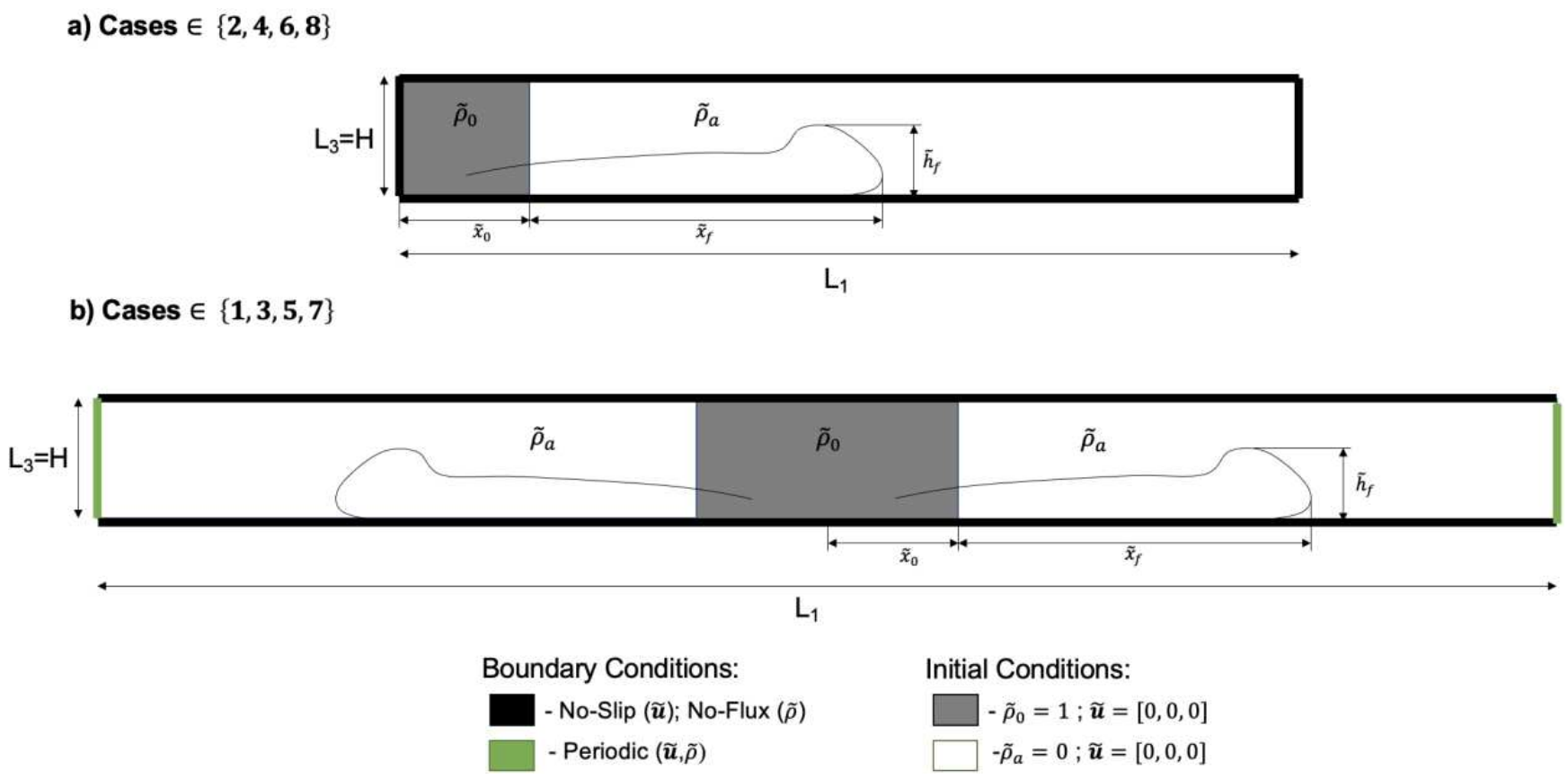

*Figure 4: Initial and boundary conditions of saline gravity current simulations of a) Cases* $\in \{2,4,6,8\}$ *and b) Cases* $\in \{1,3,5,7\}$.

The geometry and boundary conditions applied in Cases $\in \{1,3,5,7\}$, outlined in Figure 4b, are in alignment with those used by Cantero et al. (2007, 2008). A no-slip boundary condition for the velocity field and a no-flux boundary condition for the concentration field is applied on the upper and lower boundaries, while the end and span-wise wall are periodic, thereby producing a periodic array of gravity currents. The initial velocity was zero throughout the domain, while the concentration was set to $\tilde{\chi} = 1$ within the lock, and $\tilde{\chi} = 0$ elsewhere. However, due to the periodic span-wise boundaries, a small random perturbation was applied at the surface of the gate to initiate a break-down into fully three-dimensional flow.

No-slip and no-flux boundary conditions in the LBM-LES models were implemented using half-way bounce-back boundary conditions, as is conventional in the field.[43,44,52,63] Simulations in RAFSINE and VirtualFluids were run on a regularly spaced grid of nodes with a non-dimensional grid spacing of $\Delta\tilde{x} = H/N_{L_3}$. The Nek5000 simulations for Cases 2 and 4 were run on grids of resolution 1600x120x120 and 2000x184x184 respectively.

The LBM simulations are run in lattice units (LU), where conversion between LU and non-dimensional units is achieved through the length scale $\Delta\tilde{x}$ , and time scale $\Delta\tilde{t}$. Appropriate Reynolds number scaling in the LBM models is achieved by adjusting viscosity in lattice units. The characteristic velocity in lattice Boltzmann units is fixed at a value of $U_b \text{LU} = 0.03$ i.e., $\widetilde{U}_b = U_b \text{LU}\ \Delta\tilde{x}/\Delta\tilde{t} = 0.03\Delta\tilde{x}/\Delta\tilde{t}$. This was motivated by the recommendations of Krüger et al.[44] on maintaining accuracy and stability. As the value of the non-dimensional velocity scale is $\widetilde{U}_b = 1$, the non-dimensional time step for a simulation is given by $\Delta\tilde{t} = 0.03/N_{L_3}$. The reduced gravity for a simulation can then be calculated as $\tilde{g}' = g'\ \text{LU}\ \Delta\tilde{x}/\Delta\tilde{t}^2 = \frac{0.03^2}{N_{L_3}}\Delta\tilde{x}/\Delta\tilde{t}^2$. As the characteristic length scale $H = N_{L_3}\Delta\tilde{x}$, viscosity is calculated as $\tilde{\nu} = \nu\ \text{LU}\ \Delta\tilde{x}^2/\Delta\tilde{t} = \frac{N_{L_3}0.03}{Re_b}\Delta\tilde{x}^2/\Delta\tilde{t}$, for a given Reynolds number. In this study the Schmidt number is set to 1, therefore $\tilde{\nu} = \widetilde{D}$ in all simulations.

## 3. Results and Discussion

In this section the LBM-LES models are assessed based on their ability to predict the following key characteristics of a lock-exchange gravity current; front location and velocity within the slumping phase, transition to the inertial and/or viscous phases, the development of the correct qualitative turbulent flow features in the head and body, and wall shear stress on the lower boundary. Additionally, the computational expense of the RAFSINE and VirtualFluids codes is compared to recently developed NS-LES models of equivalent accuracy.

### 3.1. Slumping Phase

As outlined in Section 1, the slumping phase is characterized by a period of constant front velocity, in which $\tilde{x}_f \propto \tilde{t}$. In the numerical simulations, the span-averaged location of the current head, $\tilde{x}_f$, was determined by calculating the span-wise average of the density field, $\bar{\rho}(\tilde{x},\tilde{z}) = \frac{1}{L_2}\int_0^{L_2}\tilde{\rho}\,d\tilde{y}$, and then searching from $\tilde{x} = L_1$ to $\tilde{x} = 0$ for the first node with a density below the interface threshold of $\tilde{\rho} = 0.02$. This threshold has been used previously by Ottolenghi et al.[43] and it was easily verified that the computed front location was insensitive to variations in the threshold.

Plots of current front location against time are presented for each case in Figure 5, except for Case 7 as validation data was not readily available for a direct comparison in the Cantero et al.[21] publication. The $\tilde{x}_f \propto \tilde{t}$ curve is also plotted for reference. In each case both RAFSINE and VirtualFluids accurately capture the slumping phase, predicting a constant gradient in $\bar{x}_f$ until $\tilde{t} \approx 10$. Additionally, within the slumping phase, the computed front locations from both LBM-LES codes are in close agreement with the DNS results of Nek5000 and the Cantero et al.[18] study. In cases 6 and 8, where the codes are validated against experimental results, good agreement is observed with the experiments and LBM simulations of Ottolenghi et al.[43].

The accuracy of the slumping phase simulation can be verified more quantitively through the Froude number, defined in Equation 59 as the constant non-dimensional front velocity within the slumping phase. It is evaluated within the time range of $2.5 \leq \tilde{t} \leq 10$.

$$Fr = \left.\frac{d\bar{x}_f}{dt}\right|_{2.5 \leq \tilde{t} \leq 10} \tag{59}$$

The percentage error in the front velocity predictions of RAFSINE and VirtualFluids are presented in Table 2, where error is calculated relative to the results in the validation sources listed in Table 1, and is reported to two decimal places. All errors are less than 5%, demonstrating close quantitative agreement with the reference data.

The accuracy of the Froude number predictions of both codes is equal to that of conventional NS-LES models. Ooi et al.[64] reported Froude number predictions within $\pm 0.01$ of the reference data in their validation of a finite-volume LES code against the Hacker et al.[65] lock-exchange experimental results, for Reynolds numbers of $Re_b \in \{30\,980,\ 47\,750, 87\,750\}$.

More recently, an LES study conducted by Pelmard et al.[38], also using a finite-volume method code, observed an error of approximately 4.17%, when validating the results of a $Re_b = 60000$ simulation against the experiments of Keulegan[66].

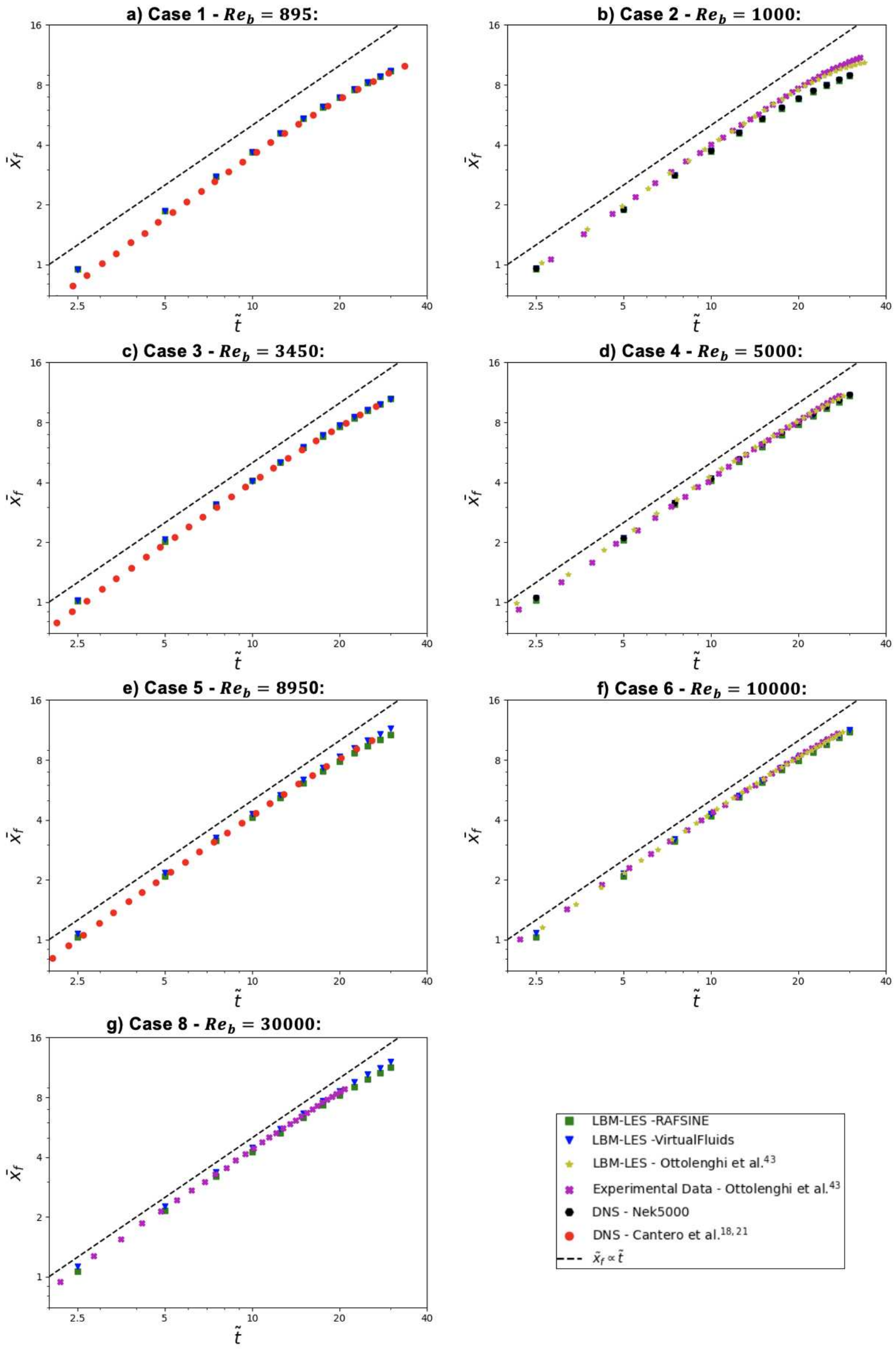
a) Case 1 - $Re_b = 895$:
b) Case 2 - $Re_b = 1000$:
c) Case 3 - $Re_b = 3450$:
d) Case 4 - $Re_b = 5000$:
e) Case 5 - $Re_b = 8950$:
f) Case 6 - $Re_b = 10000$:
g) Case 8 - $Re_b = 30000$:
$\tilde{x}_f$
$\tilde{t}$
LBM-LES -RAFSINE
LBM-LES -VirtualFluids
LBM-LES - Ottolenghi et al.[43]
Experimental Data - Ottolenghi et al.[43]
DNS - Nek5000
DNS - Cantero et al.[18,21]
$\tilde{x}_f \propto \tilde{t}$

*Figure 5: Validation of the front location predictions of RAFSINE and VirtualFluids for a) Case 1, b) Case 2, c) Case 3, d) Case 4, e) Case 5, f) Case 6, and g) Case 8. Results are validated against the DNS results from Nek5000 and those of Cantero et al.[18], as well as the experimental and numerical results of Ottolenghi et al.[43]. In h) the front location prediction from the* $Re_b = 30000$ *Pelmard et al.[34] NS-LES simulation is also plotted.*

| Case | $Re_b$ | DNS/Experimental Result | | VirtualFluids | | RAFSINE | |
|---|---|---|---|---|---|---|---|
| | | Data Source | Fr | Fr | Error (%) | Fr | Error (%) |
| 1 | 895 | DNS - Cantero et al.[18] | 0.36 | 0.37 | **2.70** | 0.36 | **0.00** |
| 2 | 1000 | DNS - Nek5000, Present Study | 0.37 | 0.37 | **0.00** | 0.37 | **0.00** |
| 3 | 3450 | DNS - Cantero et al.[18] | 0.41 | 0.41 | **0.00** | 0.40 | **2.44** |
| 4 | 5000 | DNS - Nek5000, Present Study | 0.42 | 0.42 | **0.00** | 0.41 | **2.38** |
| 5 | 8950 | DNS - Cantero et al.[18] | 0.42 | 0.43 | **2.38** | 0.41 | **2.38** |
| 6 | 10000 | Exp. - Ottolenghi et al.[43] | 0.42 | 0.42 | **0.00** | 0.42 | **0.00** |
| 7 | 15000 | DNS - Cantero et al.[21] | 0.42 | 0.44 | **4.65** | 0.42 | **0.00** |
| 8 | 30000 | Exp. - Ottolenghi et al.[43] | 0.44 | 0.45 | **2.27** | 0.43 | **2.27** |

*Table 2:Validation of the Froude number predictions of RAFSINE and VirtualFluids against DNS results and experimental data.*

In Cases 4 and 6, $Re_b \in \{5000, 10000\}$, both RAFSINE and VirtualFluids demonstrate similar accuracy to the 3D LBM model of Ottolenghi et al.[43], displaying errors of less than 3%. However in Case 2, $Re_b = 1000$, both RAFSINE and VirtualFluids predict the same Froude number as the DNS result, while the Ottolenghi et al.[43] model has an error of 5.4%. Additionally, it is clear from Figure 5b that although there is close agreement in the front location predictions of RAFSINE/VirtualFluids and the Nek5000 result, there are discrepancies between the front location predicted by the Nek5000 DNS simulation and the numerical and experimental results of Ottolenghi et al.[43]. These discrepancies are better understood through analysis of the front velocity and transition to the inertial and/or viscous phases of the flow, which is presented in Section 3.1.

In addition to achieving close quantitative agreement with the reference DNS and experimental results, visualizations of the density fields in the LBM-LES simulations demonstrates that the models capture the qualitative features of a slumping lock-exchange gravity current flow. Iso-contours of density $\tilde{\rho} = 0.02$, the chosen interface between the current and ambient, are presented in Figure 6 for cases with $Re_b \in \{1000, 5000, 10000, 30000\}$ at $\tilde{t} = 10$, to illustrate the development of turbulent flow features in the LBM-LES simulations across a range of Reynolds numbers. The iso-contours are plotted in a sub-region of the full computational domain, ranging from $\tilde{x} \in [0, 6]$, $\tilde{y} \in [0, 1]$, and $\tilde{z} \in [0, 1]$.

Density iso-contours from the RAFSINE, VirtualFluids and Nek5000 DNS simulations of Case 2 and 4, $Re_b = 1000$ and $5000$, are presented in Figure 6a and b respectively. The results show that in both cases the structure of the current interface in the LBM-LES models shows close agreement with the DNS simulation. The iso-contours of Case 2, $Re_b = 1000$, do not exhibit any significant turbulent flow features. The interface between the dense current and ambient is relatively smooth and flat in the body, and the head advances as a unified front. The current interface in Case 4, $Re_b = 5000$, exhibits clear turbulent flow features, in agreement with observations from previous DNS simulations at similar Reynolds numbers.[20] Lobe and cleft structures are visible at the lower boundary of the current front, due to instabilities at the head. Additionally, a region of turbulent mixing is evident at the current-ambient interface in the body due to the Kelvin-Helmholtz instability. It is expected that subtle differences may be observed between the interface structure at a single time (e.g. $\tilde{t} = 10$), as we are monitoring a turbulent time-dependant flow. Given this caveat, both RAFSINE and VirtualFluids show good agreement with the DNS result.

The density iso-contours produced by the RAFSINE and VirtualFluids simulations of Cases 6 and 8 are presented in Figure 6c and d respectively. Unfortunately, DNS iso-contours are not available for direct comparison. Figure 6c and d show that, as anticipated, with the increasing Reynolds number there is an intensification of the turbulent flow features observed in Case 4, Figure 6b. Whilst at low Reynolds numbers, there were negligible differences between the current-ambient interface in the RAFSINE and VirtualFluids simulations, significant differences are visible at higher Reynolds numbers.

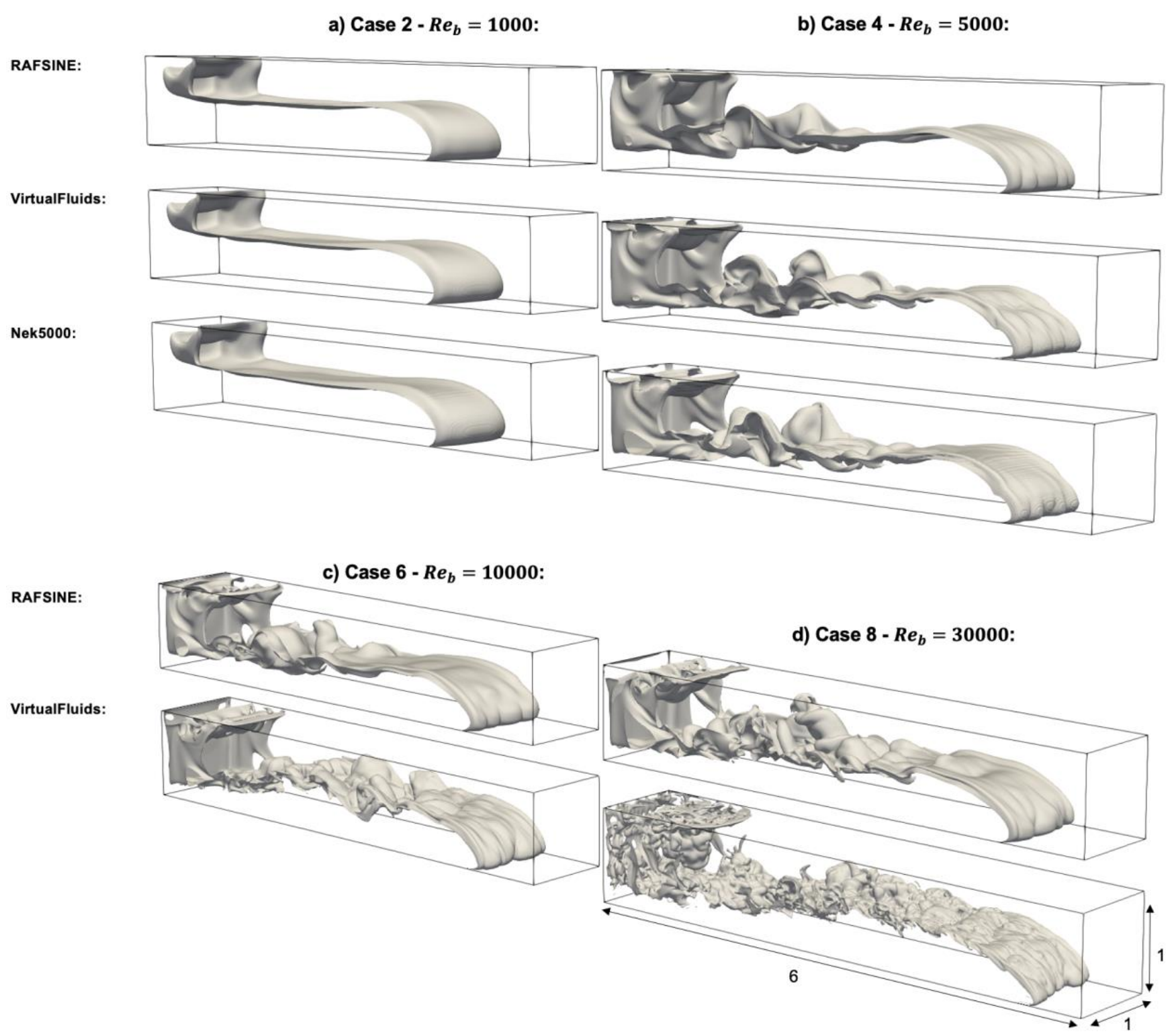

*Figure 6: Iso-contours of density $\tilde{\rho} = 0.02$ at $\tilde{t} = 10$ in the Nek5000, VirtualFluids, and RAFSINE simulations. a) Case 2, $Re_b = 1000$; b) Case 4, $Re_b = 5000$; c) Case 6, $Re_b = 10000$; d) Case 8, $Re_b = 30000$.*

The interface in the VirtualFluids simulation exhibits more small-scale folds and structures relative to the RAFSINE result, a contrast that is most clearly observed in Figure 6d. Additionally, it is anticipated that with increasing Reynolds number, the size of lobes will decrease, and their number will increase.[18,20,67] This trend is observed in the results of both LBM-LES models but appears to happen more rapidly in VirtualFluids. This is likely the result of differences in turbulence production due to shear on the lower boundary in both models. Validation of shear stress on the lower boundary, detailed in Section 3.2, indicates that shear stress is significantly underpredicted in RAFSNIE at high Reynolds numbers. As shear stress on the lower boundary is a key mechanism of turbulence generation, an underprediction of shear stress would suppress the development of turbulent flow features. Despite these observable differences in the current-ambient interface, both LBM-LES models produce equivalent accuracy in Froude number predictions across the high Reynolds number cases, see Table 1.

### 3.1. Transition to Inertial and Viscous Phases

The transition between the various phases is most clearly observed through the span-wise averaged front velocity of the current $\bar{u}_f = \frac{d\bar{x}_f}{dt}$. Plots of span-wise averaged front velocity against time are presented in Figure 7 for cases with Reynolds numbers of $Re_b \in \{1000, 5000, 8950, 10000, 15000, 30000\}$. In addition to the results from the validation data sources listed in Table 1, results are plotted from other lock-exchange experiments in the literature with relatively similar Reynolds numbers. The scaling law predictions, using the revised empirical constants determined by Cantero et al.[18], are also plotted for each case.

As outlined in Section 1, following the slumping phase the flow transitions into the inertial phase if the Reynolds number exceeds the critical value defined in Equation 8, in which case the inertial phase transition time ($\tilde{t}_{SI}$) is smaller than the viscous phase transition time ($\tilde{t}_{SV}$). It was therefore anticipated that the inertial phase would only develop in Cases 3-8.

The results presented in Figure 7 permit comparison of the LBM-LES front velocity predictions against the scaling laws, experimental data, and DNS results. Each offers a source of validation but has its own limitations. As outlined in Section 1, the scaling laws are derived directly from the governing equations but contain empirical constants that are fit by collating experimental and numerical results. Therefore, although the asymptotic behavior of the front velocity with time should match the scaling law, there may be small errors in the quantitative prediction of front velocity. Additionally, as the scaling laws were parameterized using some of the empirical data plotted in Figure 7, the scaling laws and individual experimental results are not wholly independent sources of validation.

The experimental data is a record of a real-world lock-exchange gravity current flow but is itself subject to error. Errors may accumulate from the natural variability in the system, in the measurement of material properties, the measurement of concentration and velocity fields, and in the image post-processing required to analyze the results. These limitations are revealed through the data spread when comparing experimental data sets from multiple sources with the same input parameters, as can be observed when comparing the spread of experimental data plotted in Figure 7a, Figure 7e, and Figure 7f.

DNS results provide the best source of validation data for theoretical and numerical models as the Navier-Stokes and advection-diffusion equations are solved directly across all length scales. Therefore, a DNS result forms an upper limit on simulation accuracy. However, a DNS or other numerical simulation result is still the solution of an idealized mathematical formulation of the problem, which may not map onto the physical reality and pragmatic constraints of a lock-exchange experiment.

Simulations commonly assume perfectly smooth walls, while experiments will have some degree of microscale roughness which impacts drag.[68] Additionally, the simulations assume instantaneous removal of the gate, while in reality subtle inconsistencies in the speed of gate removal may have a significant impact on the resulting current dynamics.[69,70] Nevertheless, collectively the scaling laws, experimental data, and DNS results provide a framework for assessing the accuracy of LBM-LES models.

From the relatively low Reynolds number cases ($Re_b \leq 5000$), the front velocity results for Cases 2 and 4 have been selected (Figure 7a and b), as they allow for comparisons between the results of the Nek5000 DNS simulations, the predictions of RAFSINE and VirtualFluids, and the experimental and numerical results of Ottolenghi et al.[43]. In Case 5, where $Re_b = 5000$ and $Fr = 0.42$, transition to the slumping phase occurs at $\tilde{t} \approx 11$ in the Nek5000 DNS model. The scaling laws predict that transition to the inertial phase should happen at $\tilde{t}_{SI} = 12.7$, which is in reasonable agreement with the DNS result.

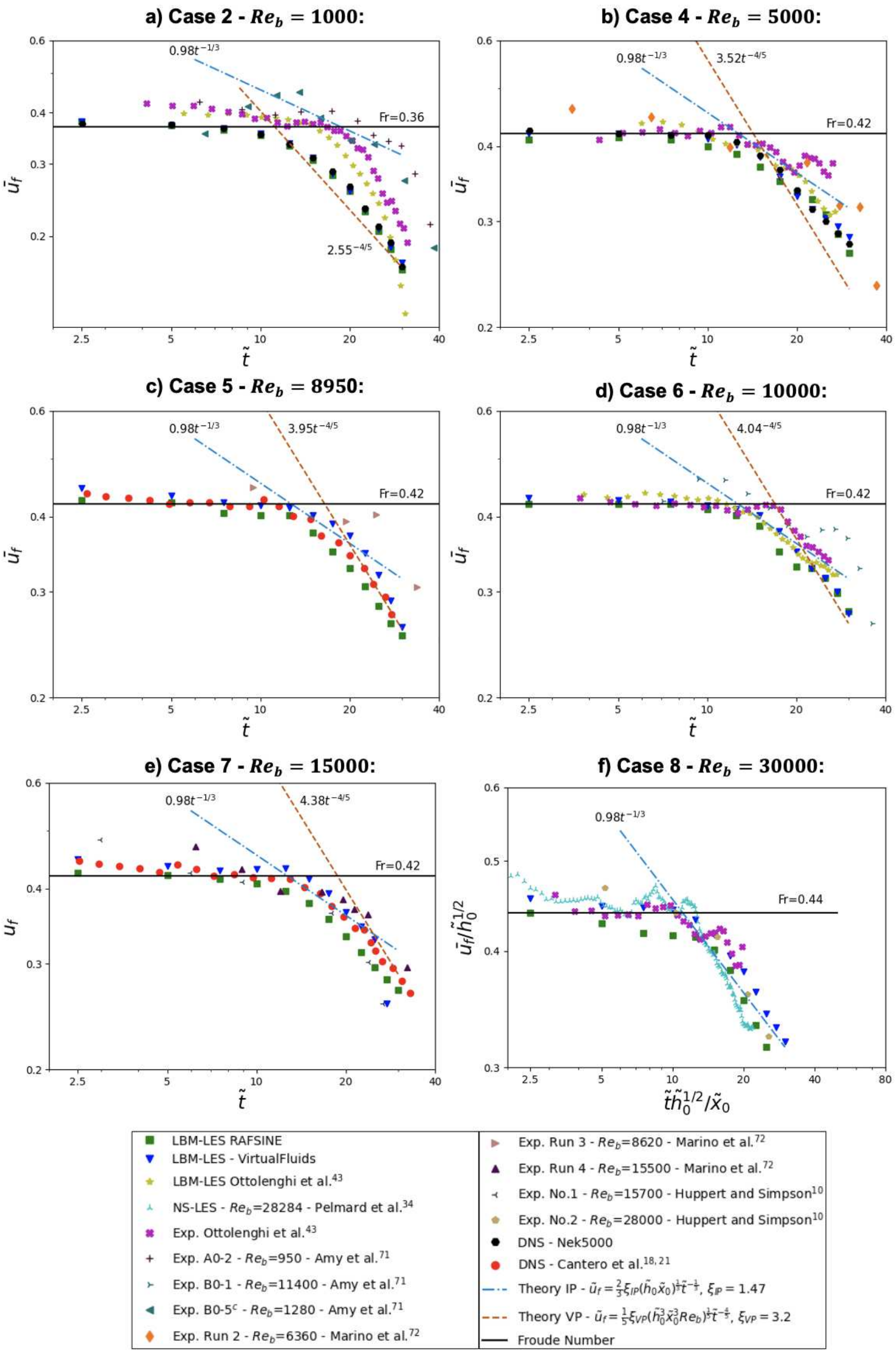
a) Case 2 - $Re_b = 1000$:
b) Case 4 - $Re_b = 5000$:
c) Case 5 - $Re_b = 8950$:
d) Case 6 - $Re_b = 10000$:
e) Case 7 - $Re_b = 15000$:
f) Case 8 - $Re_b = 30000$:
$0.98t^{-1/3}$
$2.55^{-4/5}$
$3.52t^{-4/5}$
$3.95t^{-4/5}$
$4.04^{-4/5}$
$4.38t^{-4/5}$
Fr=0.36
Fr=0.42
Fr=0.44
$\bar{u}_f$
$\tilde{t}$
$\bar{u}_f/\tilde{h}_0^{1/2}$
$\tilde{t}\tilde{h}_0^{1/2}/\tilde{x}_0$
LBM-LES RAFSINE
LBM-LES - VirtualFluids
LBM-LES Ottolenghi et al.[43]
NS-LES - $Re_b$=28284 - Pelmard et al.[34]
Exp. Ottolenghi et al.[43]
Exp. A0-2 - $Re_b$=950 - Amy et al.[71]
Exp. B0-1 - $Re_b$=11400 - Amy et al.[71]
Exp. B0-5$^c$ - $Re_b$=1280 - Amy et al.[71]
Exp. Run 2 - $Re_b$=6360 - Marino et al.[72]
Exp. Run 3 - $Re_b$=8620 - Marino et al.[72]
Exp. Run 4 - $Re_b$=15500 - Marino et al.[72]
Exp. No.1 - $Re_b$=15700 - Huppert and Simpson[10]
Exp. No.2 - $Re_b$=28000 - Huppert and Simpson[10]
DNS - Nek5000
DNS - Cantero et al.[18,21]
Theory IP - $\tilde{u}_f = \frac{2}{3}\xi_{IP}(\tilde{h}_0\tilde{x}_0)^{\frac{1}{3}}\tilde{t}^{-\frac{1}{3}}$, $\xi_{IP} = 1.47$
Theory VP - $\tilde{u}_f = \frac{1}{5}\xi_{VP}(\tilde{h}_0^3\tilde{x}_0^3 Re_b)^{\frac{1}{5}}\tilde{t}^{-\frac{4}{5}}$, $\xi_{VP} = 3.2$
Froude Number

*Figure 7: Validation of front velocity predictions of the LBM-LES models for a) Case 2, b) Case 4, c) Case 5, d) Case 6, e) Case 7, and f) Case 8. Results are validated against the DNS results from Nek5000 and those of Cantero et al.[18,21], as well as comparison to the experimental and numerical result of Ottolenghi et a.l[43]. In h) the front location prediction from the $Re_b \approx 30000$ Pelmard et al.[34] NS-LES simulation is also plotted. Front velocity within the inertial (IP) is compared to the theoretical scaling laws.[10,13–15]Within the viscous phase (VP) front velocity is compared to the theoretical scaling laws established by Huppert[17]. The scaling laws are calculated using $\xi_p = 1.47$ and $\xi_{p,Hp} = 3.2$, which are the revised values as determined by Cantero et al.[18].*

The transition time predicted by RAFSINE and VirtualFluids, as well as the experimental and numerical results of Ottolenghi et al.[43] are in good agreement with the DNS result. Within the inertial phase, front velocity in the DNS and LBM-LES simulations scale according to the $\bar{u}_f \propto \tilde{t}^{-\frac{1}{3}}$ law. The scaling laws predict a short inertial phase with the transition to the viscous phase occurring at $\tilde{t}_{IV} = 15.4$. Departure from inertial scaling at this time occurs in both the DNS and LBM-LES models, although the velocity then goes on to decay at a rate between the theoretical laws for the inertial and viscous phases. This may be due to the fact that the Reynolds number for Case 4, $Re_b = 5000$, is very close to the critical Reynolds number of $Re_{cr} = 3170$, corresponding to a case with a Froude number of $Fr = 0.42$. The critical Reynolds number is the threshold above which the current will transition from the slumping phase to the inertial phase, and below which the current will bypass the inertial phase, transitioning directly into the viscous phase. In cases where $Re_b \gg Re_{cr}$ or $Re_b \ll Re_{cr}$, closer agreement is observed with the viscous scaling laws.

For Case 2, where $Re_b = 1000$, the transition to the viscous phase occurs at $\tilde{t} \approx 10$ in the Nek5000 DNS model. The scaling laws predict a direct transition from the slumping to viscous phase at $\tilde{t}_{SV} = 11.2$, in reasonable agreement with the DNS output. Following the transition, front velocity in the DNS model scales according to the $\bar{u}_f \propto \tilde{t}^{-\frac{1}{3}}$ law as expected. The predictions of RAFSINE and VirtualFluids are in close agreement with the DNS result, both in the time of transition to the viscous phase and the subsequent decay in velocity. However, significant disparities are observed between the DNS result and the experimental and numerical predictions of Ottolenghi et al.[43]. The front velocity in the experiment declines to a constant value of $\bar{u}_f = 0.36$, before transitioning to the viscous phase at $\tilde{t} \approx 20$, much later than the transition time predicted by the scaling laws or observed in the DNS. The results of two experimental runs conducted by Amy et al.[71] for Reynolds numbers of 950 and 1280 are also plotted for reference. The Amy et al.[71] experiments also transition at $\tilde{t} \approx 20$, indicating that the Ottolenghi et al.[43] result is not an anomaly, although there is a wide spread between the experimental front velocities post transition. The 3D LBM-LES simulation of Ottolenghi et al.[43] also follows this trend. It is proposed that the difference between the DNS and experimental result is caused by greater turbulence generation in physical experiments relative to the DNS and LBM-LES simulations. Around the transitional Reynolds number, the experimental result may be more sensitive to additional sources of turbulence generation in the experiments, such as surface roughness, and disturbance triggered by the gate release, which makes the idealized DNS model a poorer representation of the experimental conditions. A review of the other cases shows that at higher Reynolds numbers ($Re_b \geq 5000$), the results of DNS simulations are in good agreement with experiments.

The front velocity results for Case 5 are presented in Figure 7c. In Case 5, $Re_b = 8950$, the front velocity predictions of RAFSINE and VirtualFluids are validated against the DNS simulations of Cantero et al.[18]. The transition times predicted by the Cantero et al.[18] DNS model are in very close agreement with the scaling laws, which predict $\tilde{t}_{SI}$ =12.7 and $\tilde{t}_{IV} = 19.8$. Additionally, the scaling laws are a very good quantitative prediction of front velocity. Both RAFSINE and Virtual fluids show good agreement with the Cantero et al.[18], correctly predicting transition times and scaling within each phase. An experimental result of Marino et al.[72], experimental Run 3 ($Re_b = 8620$), was also plotted for reference. The sparsity in front velocity readings makes direct comparison challenging, but the transition time to the inertial phase is in reasonable agreement with the Cantero et al.[18] DNS model and LBM-LES simulations, although the onset of the viscous phase appears to occur later in the Marino et al.[72] experimental run.

In Case 6 (Figure 7d.), $Re_b = 10000$, the results of RAFSINE and VirtualFluids are very similar, both showing good agreement with the scaling law predictions of $\tilde{t}_{SI}$ =12.7 and $\tilde{t}_{IV} = 19.8$. The LBM model of Ottolenghi et al.[43] shows good agreement with the scaling laws in the slumping and inertial phase, but does not transition to the viscous phase at the expected time. The experimental current of Ottolenghi et al.[43] appears to begin a transition to the inertial phase at $\tilde{t} \approx 12$, and then front velocity plateaus, before decaying according to the inertial scaling law until $\tilde{t} \approx 30$. A similar trend is observed in the experimental run B0-1 of Amy et al.[71], $Re_b = 11400$, although the onset of viscous scaling occurs at a later time due to the higher Reynolds number.

For Case 7 (Figure 7e.), $Re_b = 15000$, the front velocity predictions of RAFSINE and VirtualFluids are validated against the DNS simulation of Cantero et al.[21]. The transition times in the Cantero et al.[21] DNS results are again in very close agreement with the scaling laws, which predict transition to the inertial and viscous phases at $\tilde{t}_{SI} = 12.7$ and $\tilde{t}_{IV} = 24.7$. The LBM-LES models demonstrate good agreement with the DNS result, accurately predicting both phase transition times and scaling withing the phases. Figure 7e. also includes plots from the $Re_b = 15550$ experiment of Marino et al.[72], and the $Re_b = 15700$ experiment of Huppert and Simpson[10]. To the extent that either the RAFSINE or VirtualFluids simulation deviates from the DNS, they remain within the range of front velocities spanned by the two experimental results, indicating the models still offer a high degree of accuracy.

Case 8 (Figure 7f.), $Re_b = 30000$, is beyond the range of DNS, but the predictions of the LBM-LES models can be validated against the experiments of Huppert and Simpson[10] and Ottolenghi et al.[43]. Additionally, it is possible to compare performance against the finite-volume NS-LES model of Pelmard et al.[34], who ran a lock-exchange simulation for $Re_b = 28284$. Since the experiment of Huppert and Simpson[10] and simulation of Pelmard et al.[34] have lock-lengths of $\tilde{x}_0 > 1$, it is necessary to rescale velocity and time such that $\tilde{u}_f/(\tilde{h}_0)^{\frac{1}{2}} = u_f/(g' h_0)^{\frac{1}{2}}$ and $\tilde{t}(\tilde{h}_0)^{\frac{1}{2}}/\tilde{x}_0 = t(g' h_0)^{\frac{1}{2}}/x_0$, causing the transition to the inertial phase to collapse down to the same time regardless of lock-length, as outlined in Section 1. Both experimental results show close agreement with the scaling law prediction of $\tilde{t}_{SI}(\tilde{h}_0)^{\frac{1}{2}}/\tilde{x}_0 = 11.0$. The RAFSINE simulation transitions prematurely but shows close agreement with the experiments and scaling laws in the inertial phase. VirtualFluids and the NS-LES model of Pelmard et al.[34] transition closer to $\frac{\tilde{t}_{SI}\,(\tilde{h}_0)^{\frac{1}{2}}}{\tilde{x}_0} = 11$, but the Pelmard et al.[34] simulation decays more rapidly post transition. Both LBM-LES models displayed an equivalent degree of accuracy to the conventional NS-LES model of Pelmard et al.[34], in the prediction of front velocity and phase transition in lock-exchange gravity currents.

Insight into the internal dynamics of the gravity current can be gained from a review of the span-wise averaged density contours ($\bar{\rho}$) through time for Case 8, presented in Figure 8. The RAFSNE and VirtualFluids contours of $\bar{\rho}$ in Case 8 both display the development of spanwise coherent, Kelvin-Helmholtz instability induced, billows at early times ($\tilde{t} \leq 5$). Although these billows have been observed to undergo substantial growth in 2D simulations, it is anticipated that in a 3D simulation of a turbulent current, the billows will lose their span-wise coherence with time due to span-wise perturbations in the chaotic flow field.[18,20,21,43] This process occurs in both LBM-LES models, with span-wise coherence of the Kelvin-Helmholtz billows disintegrating in the simulations by $\tilde{t} = 10$.

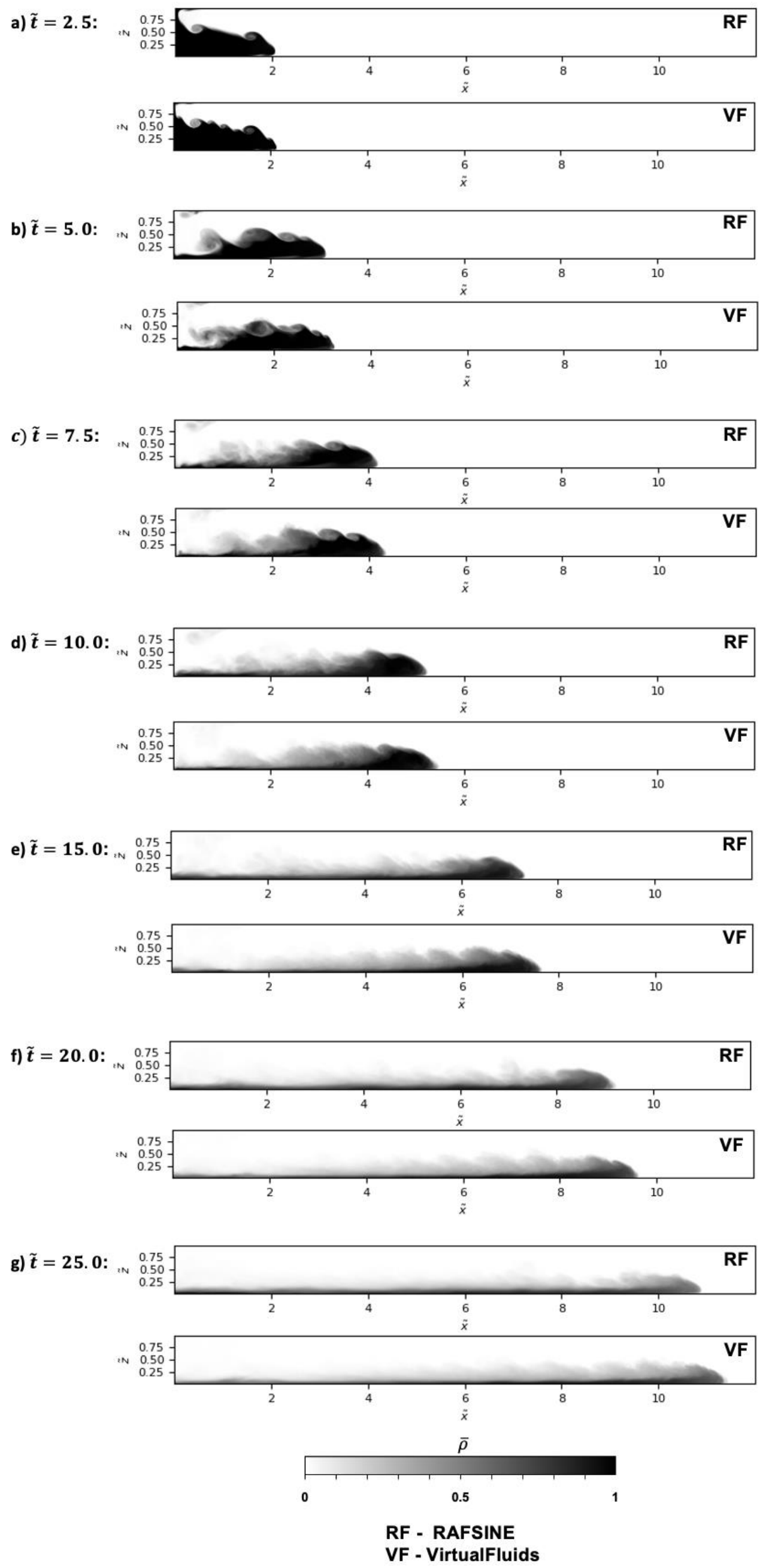
a) $\tilde{t} = 2.5$:
b) $\tilde{t} = 5.0$:
c) $\tilde{t} = 7.5$:
d) $\tilde{t} = 10.0$:
e) $\tilde{t} = 15.0$:
f) $\tilde{t} = 20.0$:
g) $\tilde{t} = 25.0$:
RF
VF
$\tilde{x}$
$\tilde{z}$
$\bar{\rho}$
0
0.5
1
RF - RAFSINE
VF - VirtualFluids

*Figure 8: Contour plots of span-wise averaged density in LBM-LES simulations of Case 8, $Re_b = 30000$, run in RAFSINE and VirtualFluids. Contours are plotted at times a) $\tilde{t} = 2.5$, b) $\tilde{t} = 5.0$, c) $\tilde{t} = 7.5$, d) $\tilde{t} = 10.0$, e) $\tilde{t} = 15.0$, f) $\tilde{t} = 20.0$, g) $\tilde{t} = 25.0$.*

### 3.2. Near-Wall Region

In this section the ability of the RAFSINE and VirtualFluids models to capture the near-wall flow characteristics of a saline gravity current are assessed by validation against DNS data from Cantero et al.[21] and the Nek5000 simulations. Accurate prediction of the flow properties in the near-wall region is necessary for simulating a number of important physical processes in environmental gravity current flows, such as turbulence production due to lower boundary shear stresses, as well as erosion and deposition of stationary material at the boundaries.

Accurate simulation of near-wall flow requires a high level of near wall resolution, which is measured by the non-dimensional distance ($z^+$) of the wall adjacent nodes from a no-slip boundary. The distance from a wall adjacent node to the boundary ($\Delta z_1$) is non-dimensionalized by friction velocity ($u_\tau$), and kinematic viscosity ($\nu$), as shown in Equation 60. The friction velocity ($u_\tau$) is defined in Equation 61, where $\tau_w$ is the wall shear stress, calculated using Equation 62.

$$z^+ = \frac{\tilde{u}_\tau \Delta \tilde{z}_1}{\tilde{\nu}} \tag{60}$$

$$\tilde{u}_\tau = \sqrt{\tilde{\tau}_w / \tilde{\rho}} \tag{61}$$

$$\tilde{\tau}_w = \tilde{\mu} \frac{\partial \tilde{u}}{\partial \tilde{z}}\bigg|_{wall} \tag{62}$$

The velocity gradient at the lower boundary $\frac{\partial \tilde{u}}{\partial \tilde{z}}\big|_{wall}$ is calculated via a second-order accurate finite difference approximation at the wall, using the fluid velocity at the two nearest fluid nodes in the $\boldsymbol{e}^g$ direction. A schematic of near-wall grid spacing in Nek5000, RAFSINE, and VirtualFluids is presented in Figure 9. The Nek5000 discretization has a wall node where $\boldsymbol{u} = 0$, and the two nearest fluid nodes in the $\hat{\boldsymbol{z}}$ direction are at spacings of $\Delta\tilde{z}_1$, and $\Delta\tilde{z}_2 > \Delta\tilde{z}_1$ as grid spacing is non-uniform in the vertical direction. In the LBM-LES models the boundary is located at a distance $\Delta\tilde{z}_1 = \Delta\tilde{x}/2$ from the nearest fluid node due to the use of the half-way bounce-back boundary condition.[44] The second fluid node is then spaced at a distance of $\Delta\tilde{x}$ from the wall adjacent node. As the Nek5000 DNS simulation has a non-uniform grid spacing in the vertical direction, and a larger number of mesh points relative to the LBM-LES models, the fluid nodes used in the Nek5000 velocity gradient approximations span a smaller $z^+$ range.

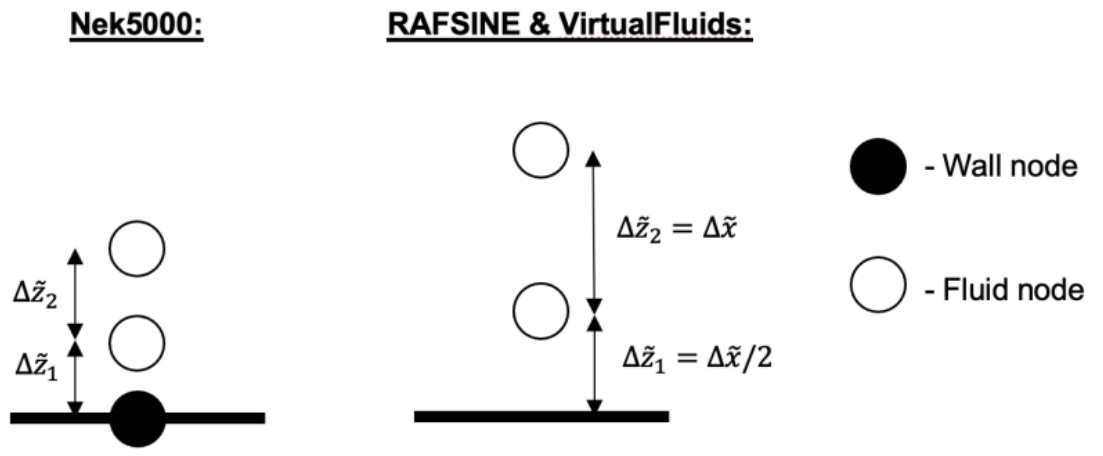

*Figure 9: Schematic of near-wall grid spacing in Nek5000, RAFSINE and VirtualFluids.*

Pelmard et al.[34], who investigated grid resolution requirements for wall resolved LES simulations of gravity currents, recommends that the maximum $z^+$of a wall adjacent node must meet the criteria $z^+ < 10$ to sufficiently capture the boundary layer. The maximum $z^+$ of wall-adjacent nodes, presented in Table *3*, shows that this standard is met in all cases. Further to this condition, the maximum $z^+$ is within the viscous sublayer of the current's boundary layer, $z^+ \leq 5$, in all cases but Cases 7 and 8 in VirtualFluids, where $z^+$ lies just within the buffer region $5 < z^+ < 30$. Overall, the range of maximum $z^+$ values is equivalent to that used by Pelmard et al.[34], who reported $2.5 < z^+_{max} < 7$ across their simulations.

| **Case** | $\boldsymbol{Re_b}$ | **Max z⁺ of Wall-Adjacent Node** | | |
|---|---|---|---|---|
| | | RAFSINE | VirtualFluids | Nek5000 |
| 1 | 895 | 0.57 | 0.60 | ~ |
| 2 | 1000 | 0.56 | 0.59 | 0.11 |
| 3 | 3450 | 1.53 | 2.10 | ~ |
| 4 | 5000 | 1.80 | 2.75 | 0.10 |
| 5 | 8950 | 2.83 | 4.45 | ~ |
| 6 | 10000 | 2.89 | 4.68 | ~ |
| 7 | 15000 | 3.83 | 5.98 | ~ |
| 8 | 30000 | 4.45 | 6.92 | ~ |

*Table 3: Maximum $z^+$ of wall adjacent node in all numerical simulations*

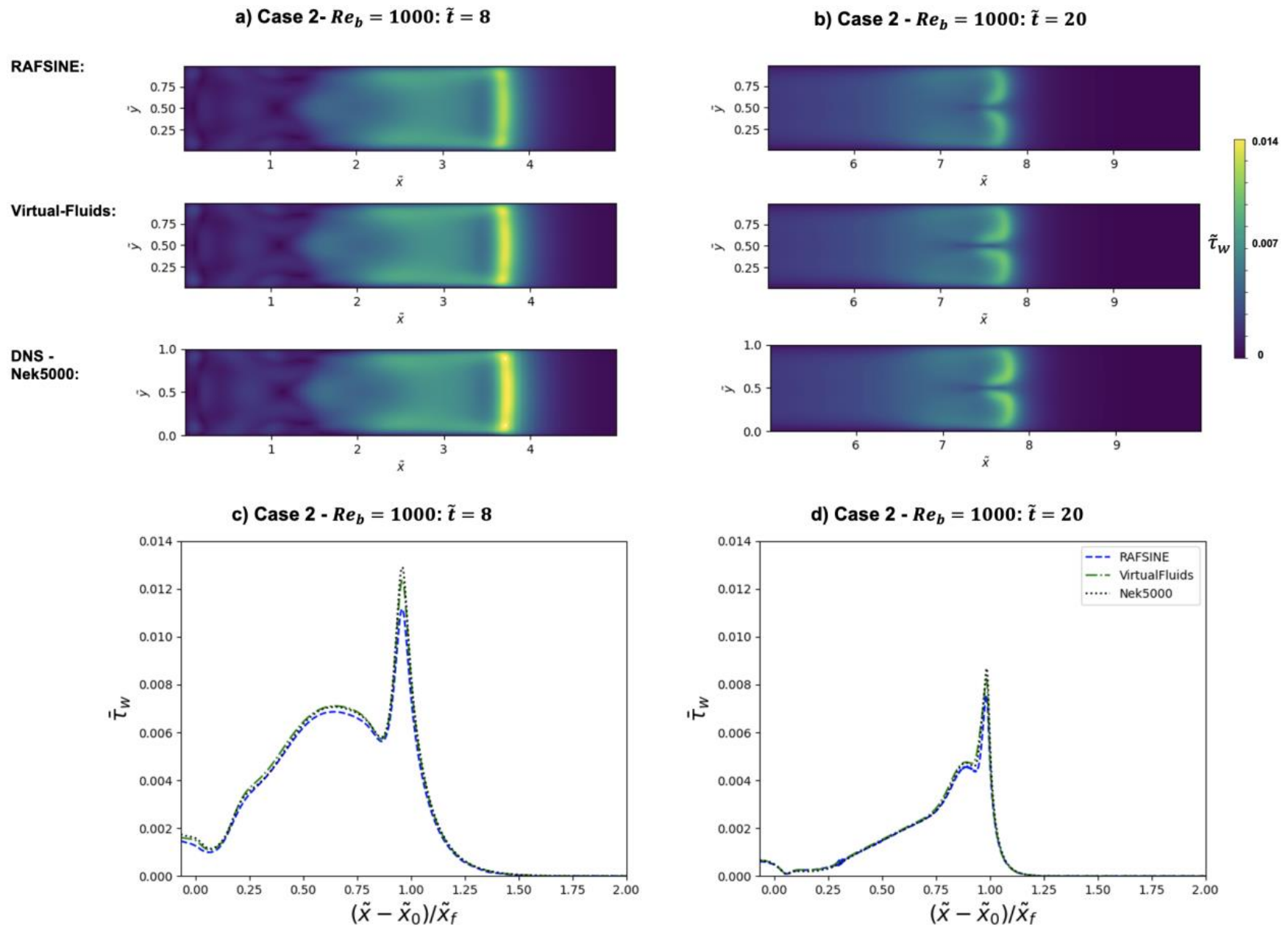

*Figure 10: Validation of predicted dimensionless shear stress on the lower boundary of the RAFSINE the VirtualFluids simulations against the Nek5000 DNS result for Case 2, $Re_b = 1000$. a) Contours of $\tilde{\tau}_w$ on the lower boundary at $\tilde{t} = 8$. b) Contours of $\tilde{\tau}_w$ on the lower boundary at $\tilde{t} = 20$. c) Plot of span-wise averaged shear stress ($\bar{\tau}_w$) at $\tilde{t} = 8$. d) Plot of span-wise averaged shear stress ($\bar{\tau}_w$) at $\tilde{t} = 20$.*

Verification of the maximum $z^+$ requires the calculation of the non-dimensional wall shear stress ($\tilde{\tau}_w$), which has been validated against DNS for Cases $\in \{2, 4, 7\}$. The validation of the $\tilde{\tau}_w$ predictions of the LBM-LES models against DNS of Case 2, $Re_b = 1000$, is presented in Figure 10, where contour plots of $\tilde{\tau}_w$ at $\tilde{t} = 8$ and $\tilde{t} = 20$ are presented in Figure 10a and Figure 10b respectively.

At $\tilde{t} = 8$ the stress pattern in the DNS result is characterized by a region of high stress along the current front, followed by a fairly uniform stress distribution in the body, and very low shear stresses in the region behind the removed gate. There is a gradual increase in stress ahead of the front due to the displacement of ambient fluid by the flow. By $\tilde{t} = 20$ two lobes have formed at the head, and the peaks in shear stress occur along the edges of the lobes. Behind the head, the spanwise stress distribution is relatively uniform but steadily decreases with distance from the head. Stresses ahead of the front are lower at $\tilde{t} = 20$ as fluid is displaced less rapidly, due to the deacceleration of the current in the viscous phase. The stress distributions predicted by RAFSINE and VirtualFluids are both in very close agreement with the DNS contours.

A more quantitative validation can be conducted by comparing plots of span-wise averaged wall shear stress $\bar{\tau}_w = \frac{1}{\tilde{L}_2}\int_0^{\tilde{L}_2} \tilde{\tau}_w d\tilde{y}$, presented in Figure 10c and d for $\tilde{t} = 8$ and $\tilde{t} = 20$ respectively. Shear stress is plotted against the rescaled streamwise distance $(\tilde{x} - \tilde{x}_0)/\tilde{x}_f$, where $\tilde{x}_f$ is the front location predicted by a given numerical model at $\tilde{t} \in \{8, 20\}$. This has the effect of collapsing the location of the front $(\tilde{x} - \tilde{x}_0 = \tilde{x}_f)$ to $(\tilde{x} - \tilde{x}_0)/\tilde{x}_f = 1$ on the graphs, which filters out errors in predicted front location in the LBM-LES models, allowing the stream-wise profiles of $\bar{\tau}_w$ to be compared relative to the front of each current.

The $\bar{\tau}_w$ distribution in the DNS result exhibits a sharp spike in shear stress at the current front, followed by a smaller rounded peak in the body, after which stress decreases with distance from the head. Profiles of $\bar{\tau}_w$ in RAFSINE and VirtualFluids show close quantitative agreement with the DNS result in the locations of peak $\bar{\tau}_w$, although VirtualFluids appears to perform better in predicting the magnitude of the peak stress. This is reflected in the $e_{L_1}$error in the total frictional force $(F_w)$ applied to the lower boundary, defined in Equations 63-64, where $\tilde{t}_i \in \{\tilde{t}_1, \tilde{t}_2, \dots, \tilde{t}_{N_t}\}$ is a list of $N_t = 12$ times at which a results file was output.

$$e_{L_1} = \frac{1}{N_t}\sum_i \left|\frac{F_w^{DNS}(\tilde{t}_i) - F_w^{LBM-LES}(\tilde{t}_i)}{F_w^{DNS}(\tilde{t}_i)}\right| \quad (63)$$

$$F_w(\tilde{x}, \tilde{y}, \tilde{t}_i) = \int_0^{\tilde{L}_2}\int_0^{\tilde{L}_1} \tilde{\tau}_w(\tilde{x}, \tilde{y}, \tilde{t}_i)\,d\tilde{x}\,d\tilde{y} \quad (64)$$

In Case 2, the $e_{L_1}$ error in RAFSINE and VirtualFluids was 6.9% and 1.9% respectively, demonstrating that although both models show good agreement with the DNS result, the error in the VirtualFluids prediction is less than a third of the RAFSINE model error.

Contours of $\tilde{\tau}_w$ in the RAFSINE, VirtualFluids and Nek5000 simulations of Case 4 are presented in Figure 11a and b at $\tilde{t} = 8$ and $\tilde{t} = 20$ respectively. At $\tilde{t} = 8$ the DNS contours display peaks in shear stress at the lobes of the current front, followed by a roughly circular region of high stress within the body. This secondary peak in stress is less evident at $\tilde{t} = 20$, where the high stress regions are concentrated at the head. The qualitative stress distribution is well reproduced in RAFSINE and VirtualFluids, as both display high stresses at the current front, and a second circular high stress region in the body at $\tilde{t} = 8$. However, the contours indicate that the peaks in stress are lower in the RAFSINE model than in the DNS result. This is confirmed by the plots of span-wise averaged shear stress in Figure 11c and d. Although the qualitative structure of the $\bar{\tau}_w$ profile is captured by RAFSINE, the magnitude of the peaks is substantially lower than the DNS result. The magnitude of the peak stress in the VirtualFluids result is in close agreement with the DNS at $\tilde{t} = 8$, but is underpredicted at $\tilde{t} = 20$. This is evident in the $e_{L_1}$ errors for Case 4, where the VirtualFluids model has an error of 4.9%, still in good agreement with DNS, while the RAFSINE model error was 33.4%. This reflects a known limitation of the standard Smagorinsky turbulence model in the near wall region when simulating turbulent flows.[34,36] As $C_s$ is held constant, see Section 2.3.1, the eddy-viscosity may be non-zero in the near-wall region, which reduces the velocity gradient at the wall, thereby decreasing $\tilde{\tau}_w$ and artificially increasing the thickness of the boundary layer. Although this can be overcome through the use of Van Driest style damping or a dynamic Smagorinsky model, researchers often use the standard Smagorinsky model when concerned with capturing flow features far from the wall.[34,36,38,43] As outlined in Section 2.3.1, the VirtualFluids LBM formulation does not include a sub-grid eddy-viscosity turbulence model, and so is not impacted by this limitation.

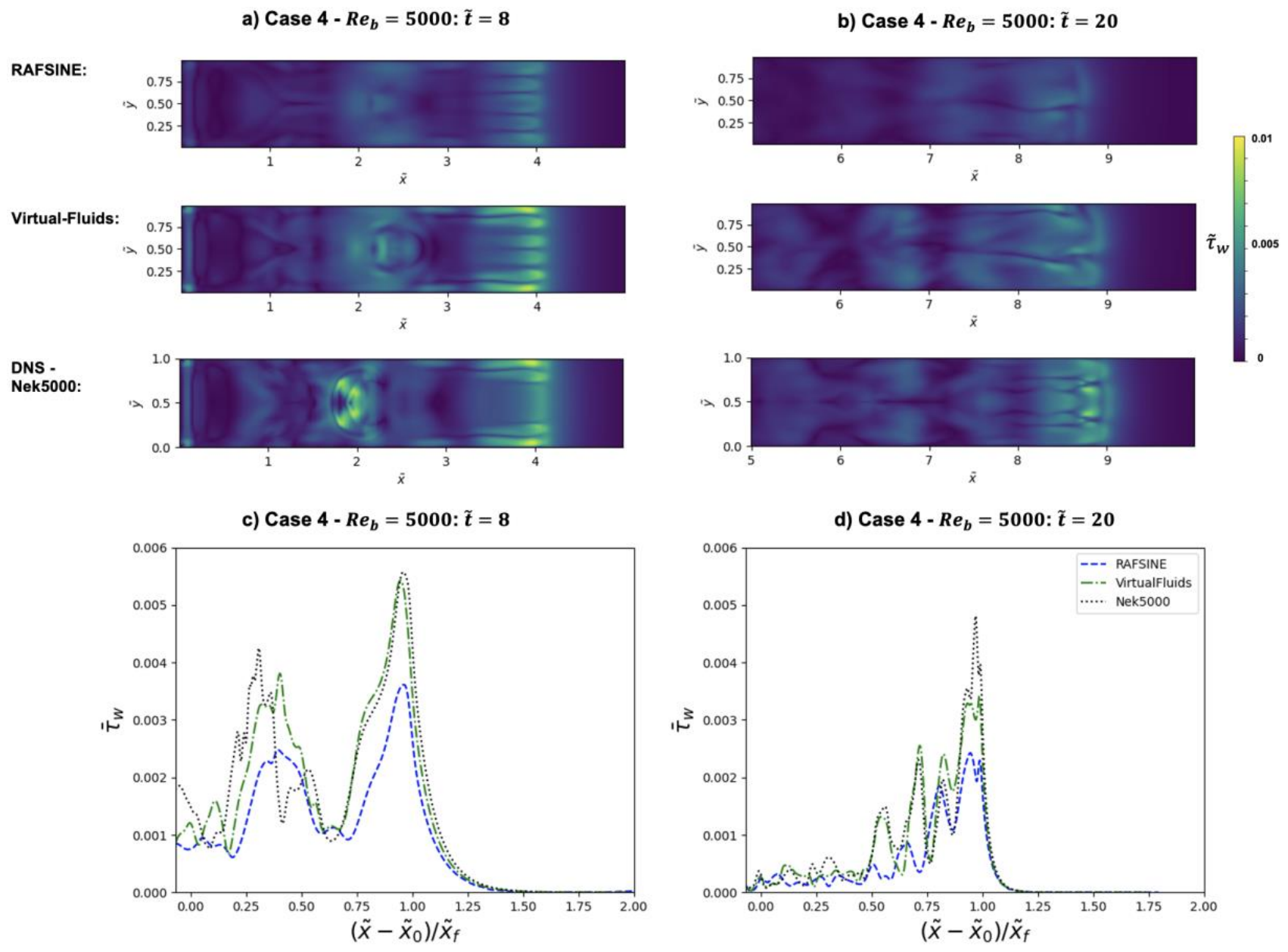

*Figure 11: Validation of predicted dimensionless shear stress on the lower boundary of the RAFSINE the VirtualFluids simulations against the Nek5000 DNS result for Case 4, $Re_b = 5000$. a) Contours of $\tilde{\tau}_w$ on the lower boundary at $\tilde{t} = 8$. b) Contours of $\tilde{\tau}_w$ on the lower boundary at $\tilde{t} = 20$. c) Plot of span-wise averaged shear stress ($\bar{\tau}_w$) at $\tilde{t} = 8$. d) Plot of span-wise averaged shear stress ($\bar{\tau}_w$) at $\tilde{t} = 20$.*

Similar trends are observed in Case 7 (Figure 12). In this case the contour plots of $\tilde{\tau}_w$, Figure 12a and b, are compared to those produced by Cantero et al.[21], which unfortunately were published without a color scale, making direct comparison challenging. However, the streaking stress patterns in the head, clearly observable in the DNS result, are reproduced in both RAFSINE and VirtualFluids, as well as a banded region of high stress in the body at $\tilde{t} = 8$. The stress distribution in RAFSINE is noticeably smoother than that observed in the Cantero et al.[21] DNS result and VirtualFluids, which is due to the previously discussed spurious damping in the near-wall region caused by the standard Smagorinsky model. Plots of $\bar{\tau}_w$, Figure 12c and d, show that RAFSINE substantially underpredicts shear stresses, while the VirtualFluids profile shows reasonable quantitative agreement, although still marginally underpredicting the frontal peak in stress. As Cantero et al.[21] does not provide quantitative stress data for the lower boundary, but does provide plots of $\bar{\tau}_w$ at $\tilde{t} = 8$ and $\tilde{t} = 20$, the $e_{L_1}$ error is calculated using the spanwise averaged frictional force on the wall ($\bar{F}_w$), as shown in Equations 65-66, where $\tilde{t}_i \in \{8, 20\}$. Using this revised definition, the $e_{L_1}$ error in RAFSINE becomes substantial at 50.6%, while the VirtualFluids model remains in good agreement with DNS at 8.8%.

$$e_{L_1} = \frac{1}{N_t}\sum_i \left|\frac{\bar{F}_w^{DNS}(\tilde{t}_i) - \bar{F}_w^{LBM-LES}(\tilde{t}_i)}{\bar{F}_w^{DNS}(\tilde{t}_i)}\right| \tag{65}$$

$$\bar{F}_w(\tilde{x}, \tilde{t}_i) = \int_0^{\tilde{L}_1} \bar{\tau}_w(\tilde{x}, \tilde{t}_i)\, d\tilde{x} \tag{66}$$

In a modelling scenario where close quantitative agreement with DNS is essential, errors could be reduced through the application of hierarchical grids to increase resolution in the near wall region.[57]

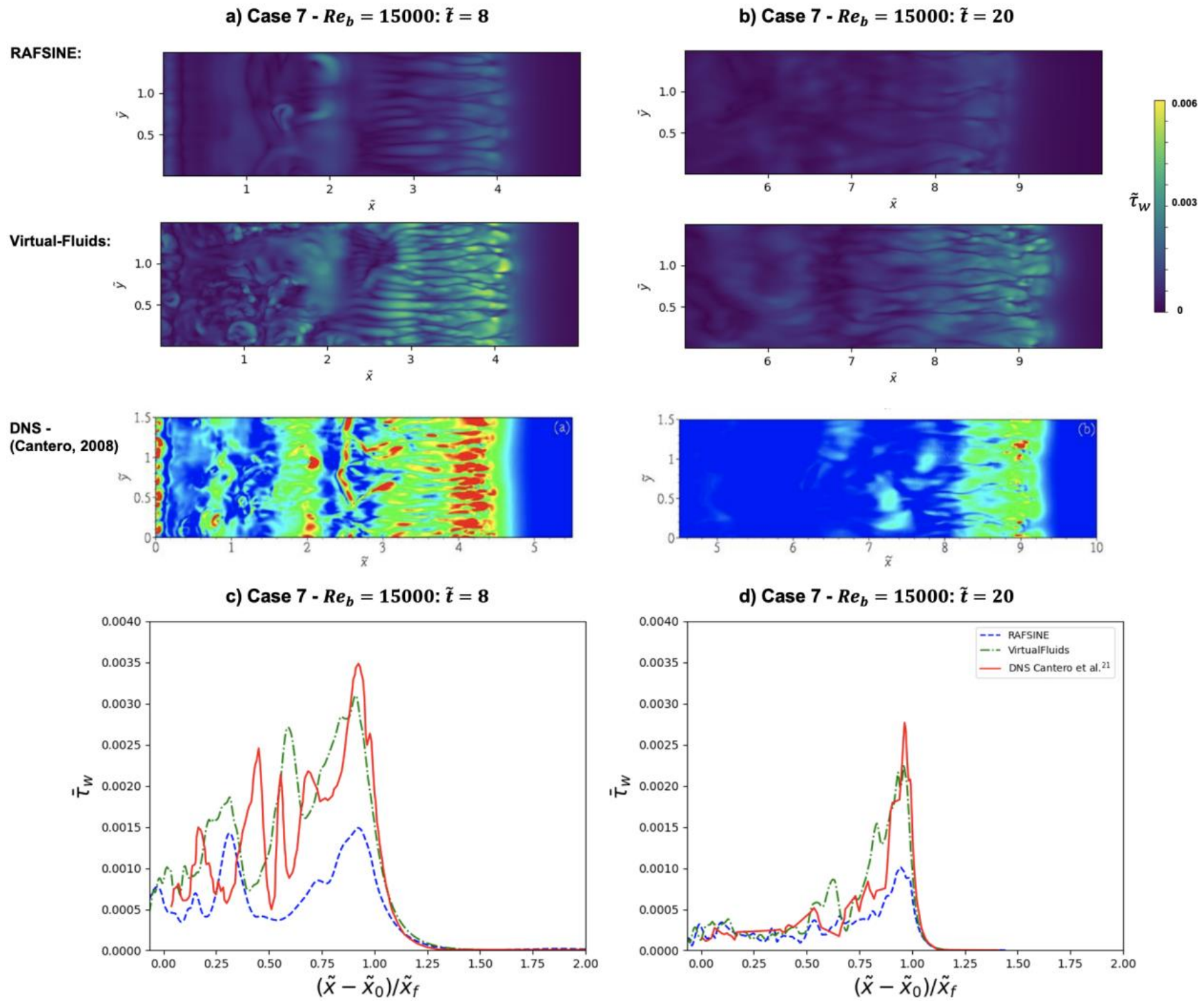

*Figure 12: Validation of predicted dimensionless shear stress on the lower boundary of the RAFSINE the VirtualFluids simulations against Cantero et al.[21] result for Case 7, $Re_b = 15000$. a) Contours of $\tilde{\tau}_w$ on the lower boundary at $\tilde{t} = 8$. b) Contours of $\tilde{\tau}_w$ on the lower boundary at $\tilde{t} = 20$. c) Plot of span-wise averaged shear stress $(\bar{\tau}_w)$ at $\tilde{t} = 8$. d) Plot of span-wise averaged shear stress $(\bar{\tau}_w)$ at $\tilde{t} = 20$.*

### 3.3. Computational Performance of the LBM-GPU Framework

As outlined in Section 2.3, the LBM-LES codes are accelerated by exporting computations to a GPU at each time step, rather than completing the tasks on the CPU. The core motivation for shifting to this LBM-GPU framework is that GPU acceleration reduces the elapsed time of a simulation relative to an equivalent implementation that runs exclusively on CPUs. However, direct comparison between the computational performance of CPU and GPU implementations is complicated by the fact that they run on different hardware using different data structures, and to the authors' knowledge consensus on a suitable metric has not been established.

When comparing the performance of two numerical models implemented to run across CPU cores, the total CPU time ($T_{CPU}$) of each program, i.e. the total time taken to process instructions, would be used to measure the computational cost of the numerical models. The relationship between $T_{CPU} = \sum_{i,j} t_{P_j}^{i}$ and elapsed time is illustrated in Figure 13a, where $t_{P_j}^{i}$ is a single block of processing time on one of $n$ processors $P_j \in \{P_1, P_2, \ldots, P_n\}$. It seems appropriate to extend this metric of total CPU time to comparisons between conventional NS models written to run on CPUs and LBM implementations in which the CPU exports computations to a GPU.

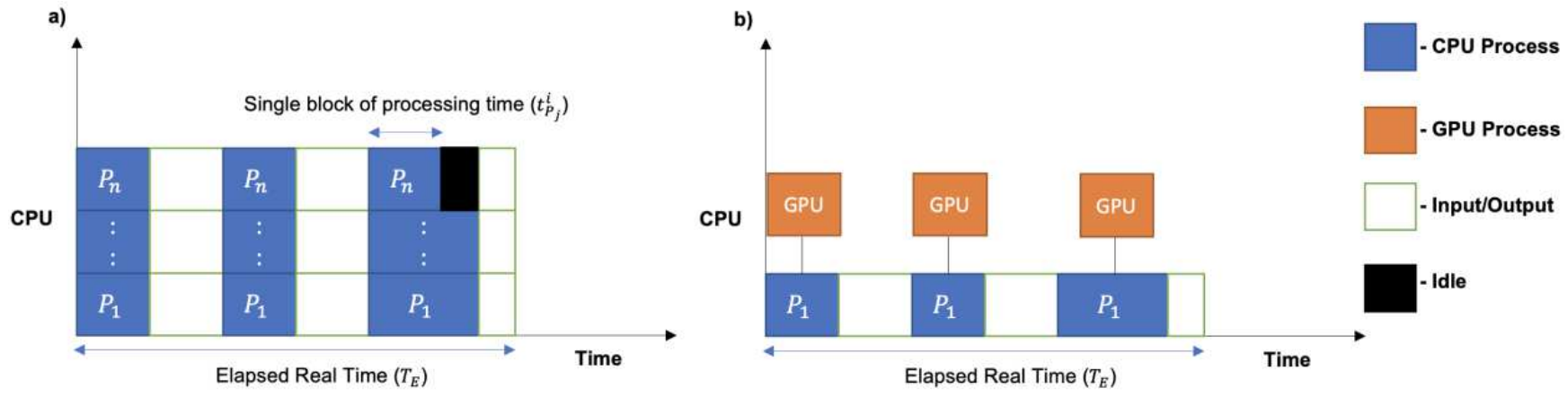

$$T_{CPU} = \sum_{i,j} t_{P_j}^{i} \qquad T_{CPU} > T_E$$

*Figure 13: Schematic of total CPU time as it relates to a) CPU implementations, and b) GPU accelerated implementations.*

In both RAFSINE and VirtualFluids, total CPU time is approximately equal to elapsed time, as the programs are executed on a single CPU core that exports data to the GPU device, as illustrated in Figure 13b. Therefore, observed speedups in total CPU time as a result of the GPU acceleration will not translate directly into equivalent speed-ups in elapsed time relative to a NS model running exclusively on CPU cores. The realized reduction in elapsed time will depend upon the number of cores used to run the CPU code, the efficiency of the parallel implementation, and the time taken for input/output operations.

Using this framework, the speedup offered by the GPU-accelerated LBM-LES codes relative to DNS in Nek5000 has been evaluated by comparing both total CPU time and elapsed time at low Reynolds numbers ($Re_b \leq 5000$). The Nek5000 simulations were run on ARC4, part of the High-Performance Computing facilities at the University of Leeds UK. Compute nodes on the cluster contain two Intel Xeon Gold 6138 CPUs ('Sky Lake'), each with 20 cores, a clock rate for non-AVX instructions of 2.0GHz, and are connected with Infiniband EDR of 100Gbit/s. The $Re_b = 1000$ and $Re_b = 5000$ Nek5000 simulations were run across 100 and 250 cores respectively. In the LBM-LES codes computations were exported to an NVIDIA V100 Tensor Core GPU.

The speedups in $T_{CPU}$ and $T_E$ are presented in Table 4, where times are written in the format hrs:mins:sec. Both RAFSINE and VirtualFluids reduce total processing time relative to Nek5000 by a factor of $10^2$ for the $Re_b = 1000$ simulation, and $10^{2.6}$ for the $Re_b = 5000$ case. This translates to a reduction in elapsed time of $10^{1.2}$ and $10^2$ for the respective cases, demonstrating significant speedups relative to DNS can be achieved whilst preserving accuracy in the prediction of key flow properties.

| | | Nek5000 | | RAFSINE | | | VirtualFluids | | |
|---|---|---|---|---|---|---|---|---|---|
| | | | | | Speedup | | | Speedup | |
| Case | $Re_b$ | $T_{CPU}$ | $T_E$ | $T_{CPU}(\approx T_E)$ | $T_{CPU}$ | $T_E$ | $T_{CPU}(\approx T_E)$ | $T_{CPU}$ | $T_E$ |
| 2 | 1000 | 39:21:13 | 07:53:47 | 00:23:48 | $\mathbf{10^{2.0}}$ | $\mathbf{10^{1.3}}$ | 00:26:52 | $\mathbf{10^{1.9}}$ | $\mathbf{10^{1.2}}$ |
| 4 | 5000 | 173:27:33 | 43:29:44 | 00:25:05 | $\mathbf{10^{2.6}}$ | $\mathbf{10^{2.0}}$ | 00:27:00 | $\mathbf{10^{2.6}}$ | $\mathbf{10^{2.0}}$ |

*Table 4: Speedup in total CPU time and elapsed time of LBM-LES models relative to DNS in Nek5000. Times are presented in the format hrs:mins:sec.*

A comparison is also made to the computational cost of the NS-LES simulations of Pelmard et al.[34], who used the structured non-staggered finite-volume code described by Norris[73] and which has been applied to a wide range of problem types.[74–76] Pelmard et al.[34] ran simulations on a computing cluster at New-Zealand eScience Infrastructure (NeSI) consortium, consisting of nodes with two Intel Xeon E5-2680 Sandy Bridge 2.70 GHz CPUs, each with 8 cores. Pelmard et al.[34] report the total CPU time required to run a turbulent lock-exchange gravity current simulation, with a standard Smagorinsky sub-grid turbulence model, across a range of grid sizes. Simulations on the largest meshes were run across 128 cores, while 64 cores were used for the smaller grids. Their results are taken to be representative of the typical cost associated with modern NS-LES codes.

The total CPU time required to simulate a unit of non-dimensional time $\left(\hat{t} = \tilde{t}(\,\tilde{h}_0)^{\frac{1}{2}}/\tilde{x}_0 = t(g'h_0)^{\frac{1}{2}}/x_0\right)$ in RAFSINE, VirtualFluids, and the Pelmard et al.[34] study is presented in Figure 14. Total CPU times for the LBM-LES models are presented for grid sizes of $N_{nodes} \in \{15, 37.5, 42.2\}\text{x}10^6$, as this represents the full range of grid sizes used in the present study. The comparison indicates that the LBM-GPU framework reduces total processing time by a factor of approximately $10^3$, when contrasted with the representative total CPU time of the finite-volume NS-LES framework. Assuming a speedup due to multi-core CPU parallelization similar to that achieved by Nek5000, it is anticipated that this would translate into a reduction in $T_E$ by a factor of approximately $10^2$, when contrasted with the Pelmard et al.[34] code running across 128 cores on the NeSI cluster. This conclusion is in agreement with the findings of a similar analysis conducted by King et al.[77]. Additionally, the LBM-GPU framework offers substantial performance advantages relative to the CPU implementation of Ottolenghi et al.[43], which has enabled 3D simulations at higher Reynolds numbers.

The results in Table 4 and Figure 14 show that both RAFSINE and VirtualFluids achieve similar speedups. The standard metric for comparing the performance of two LBM-GPU codes is the number of node updates per second, typically reported in the millions i.e., MNUPS. When running on an NVIDIA V100 GPU, RAFSINE and VirtualFluids report average update rates of 1307 MNUPS and 1064 MNUPS respectively. RAFSINE has been robustly optimized by Delbosc et al.[49] to run simulations in rectilinear domains, such as a lock-exchange channel. However, a direct comparison of the efficiency of the GPU implementations cannot be made since VirtualFluids uses two $D3Q27$ lattices for the $f_{ijk}$ and $\Phi_{ijk}$ distributions, which demands more memory resources and computations than the $D3Q19$ and $D3Q6$ lattices used by RAFSINE. Additionally, the cumulant collision kernel requires more computations per time step than the BGK collision kernel. The marginal performance gap could be narrowed by the integration of the cumulant and FCM kernels, as this would eliminate redundant memory accesses caused by reading and writing distribution functions to calculate the same macroscopic variables in each kernel.

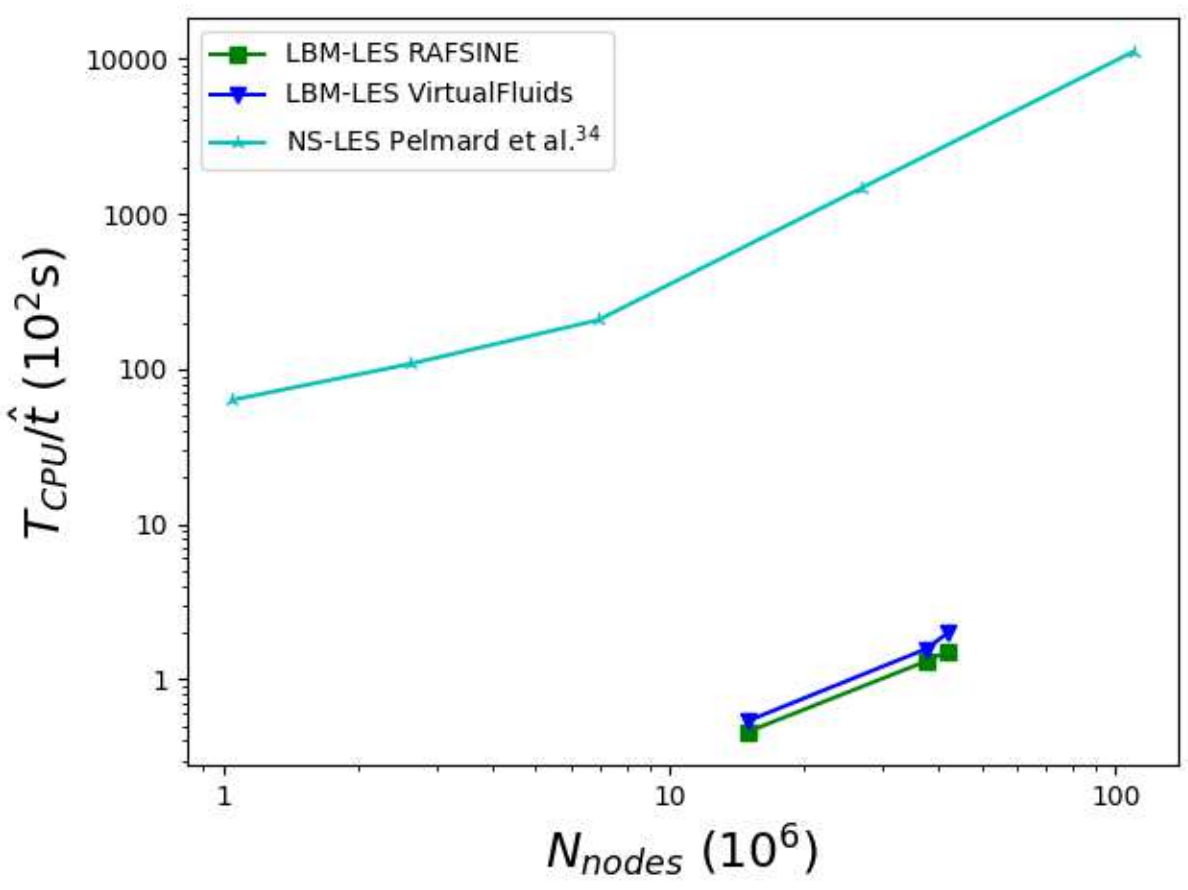

*Figure 14: Total CPU time per unit of non-dimensional time $\hat{t}$ in RAFSNIE, VirtualFluids and finite-volume LES simulations of Pelmard et al.[34].*

## 4. Conclusions

In the present study, two LBM-LES models of lock-exchange gravity currents are validated against high resolution simulations and experiments, with regard to their ability to capture key qualitative and quantitative features of a lock-exchange gravity current flow across a wide range of Reynolds numbers. The two codes, RAFSINE and VirtualFluids, demonstrate equivalent accuracy to conventional NS-LES solvers in predictions of front velocity in the slumping, inertial, and viscous phases of the flow. Additionally, the VirtualFluids model achieved good agreement with DNS in the prediction of shear stress on the lower boundary.

The computational performance of the LBM-GPU framework was assessed relative to the computational cost of DNS run in Nek5000, and that reported for the finite-volume NS-LES simulations of Pelmard et al.[34]. It is demonstrated that the LBM-GPU framework delivers speedups of at least one order of magnitude in the elapsed real time of a simulation relative to DNS at low Reynolds numbers ($Re \leq 5000$), and speedups of three orders of magnitude in total CPU time relative to a NS-LES model across a range of grid sizes for a fully turbulent flow.

Given the speed-up that can typically be achieved through multi-core CPU parallelization of CFD codes, it is estimated that the LBM-GPU models reduce the elapsed time required for a simulation by two orders of magnitude, whilst demonstrating equivalent accuracy. As a result, the numerical modelling framework presented herein can be used as a foundation for the development of models that capture more of the complexity of gravity currents, such as the two-way coupling between the hydrodynamics of environmental scale flows and the morphodynamics of boundaries in channels with complex geometries. This modelling objective would otherwise be too challenging to attempt due to the computational expense of conventional NS-LES codes.

## Acknowledgments

This work is supported by the Engineering and Physical Sciences Research Council (EPSRC) Centre for Doctoral Training in Fluid Dynamics grant EP/L01615X/1 for the University of Leeds; and the Turbidites Research Group consortium (AkerBP, CNOOC, ConocoPhillips, Murphy, OMV, and Oxy). The funders had no role in study design, the collection, analysis, and interpretation of data, or in the preparation of the article and decision to publish. This work was undertaken on ARC4, part of the High-Performance Computing facilities at the University of Leeds, UK.

## Author Declarations

The authors have no conflicts to disclose.

## Data Availability

The data that supports the findings of this study are available within the article.

## Software Availability

**Software Name:** RAFSINE
**Developer**: Nicolas Delbosc, Damilola Adekanye, Amirul Khan
**Year first official release:** 2014[49]
**Hardware requirements:** NVIDIA GPU with minimum compute capability 3.0
**System/Software requirements:** CMake (minimum version 3.15), C++ compiler with C++14 support, minimum CUDA Version 9.0, Paraview
**Program language:** C++, CUDA
**Availability:**
GitHub repository: https://github.com/scda-FluidsCDT/Saline_Gravity_Current_Models/tree/main/Straight_Channel_Single_Release_Lock_Exchange/RAFSINE
**Documentation:** readme.md in GitHub repository

**Software Name:** VirtualFluids
**Developer**: Institute for Computational Modelling in Civil Engineering (iRMB)
**Year first official release:** First GPU implementation in 2008[78,79]
**Hardware requirements:** NVIDIA GPU with minimum compute capability 3.0
**System/Software requirements:** CMake (minimum version 3.15), C++ compiler with C++14 support, minimum CUDA Version 9.0, Paraview
**Program language:** C++, CUDA
**Availability:**
GitLab repository: https://git.rz.tu-bs.de/irmb/virtualfluids
**Documentation:** readme.md in GitHub repository

**Software Name:** Nek5000-v19.0
**Developer**: Argonne National Laboratory
**Year first official release:** 2017
**Hardware requirements:** See documentation at https://nek5000.mcs.anl.gov/
**System/Software requirements:** See documentation at https://nek5000.mcs.anl.gov/
**Program language:** FORTRAN
**Availability:**
GitHub repository: https://github.com/scda-FluidsCDT/Saline_Gravity_Current_Models/tree/main/Straight_Channel_Single_Release_Lock_Exchange/Nek5000
**Documentation:** readme.md in GitHub repository

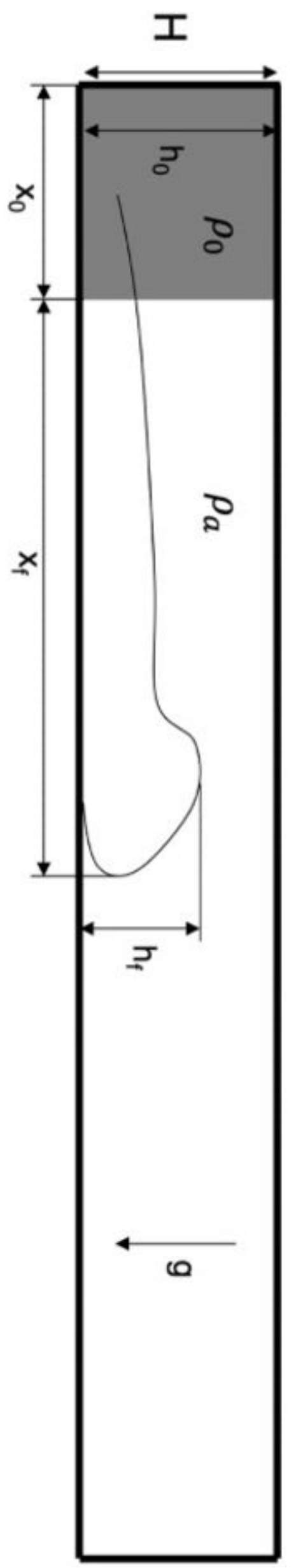
H
$h_0$
$\rho_0$
$\rho_a$
$h_f$
g
$x_0$
$x_f$

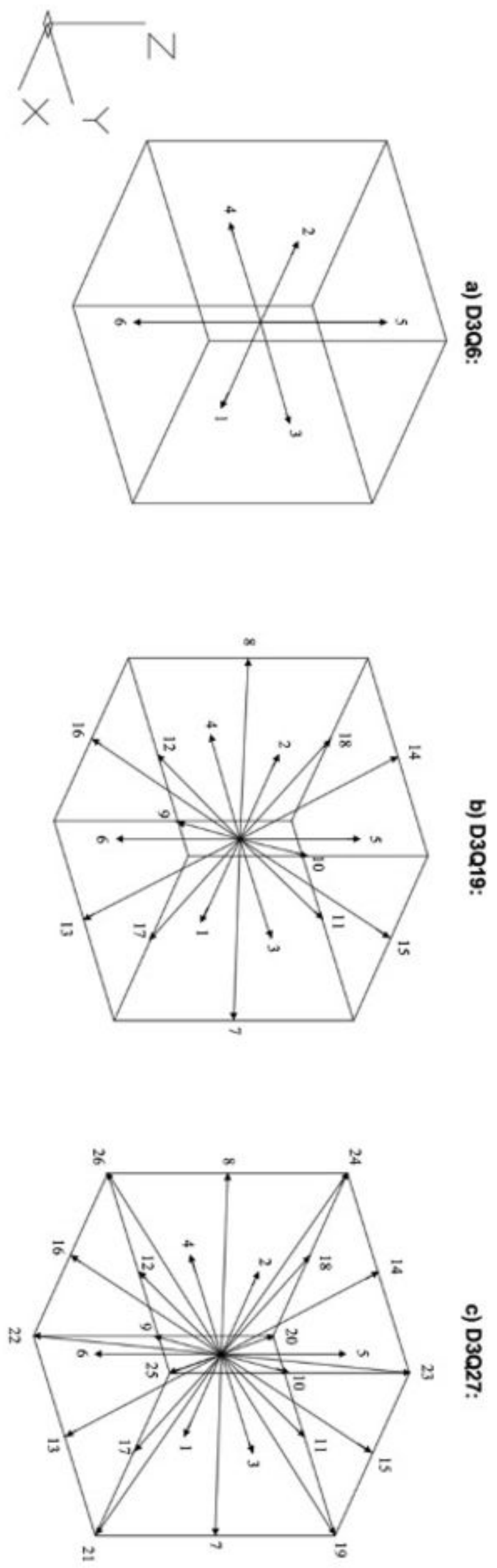
a) D3Q6:
b) D3Q19:
c) D3Q27:
X
Y
Z

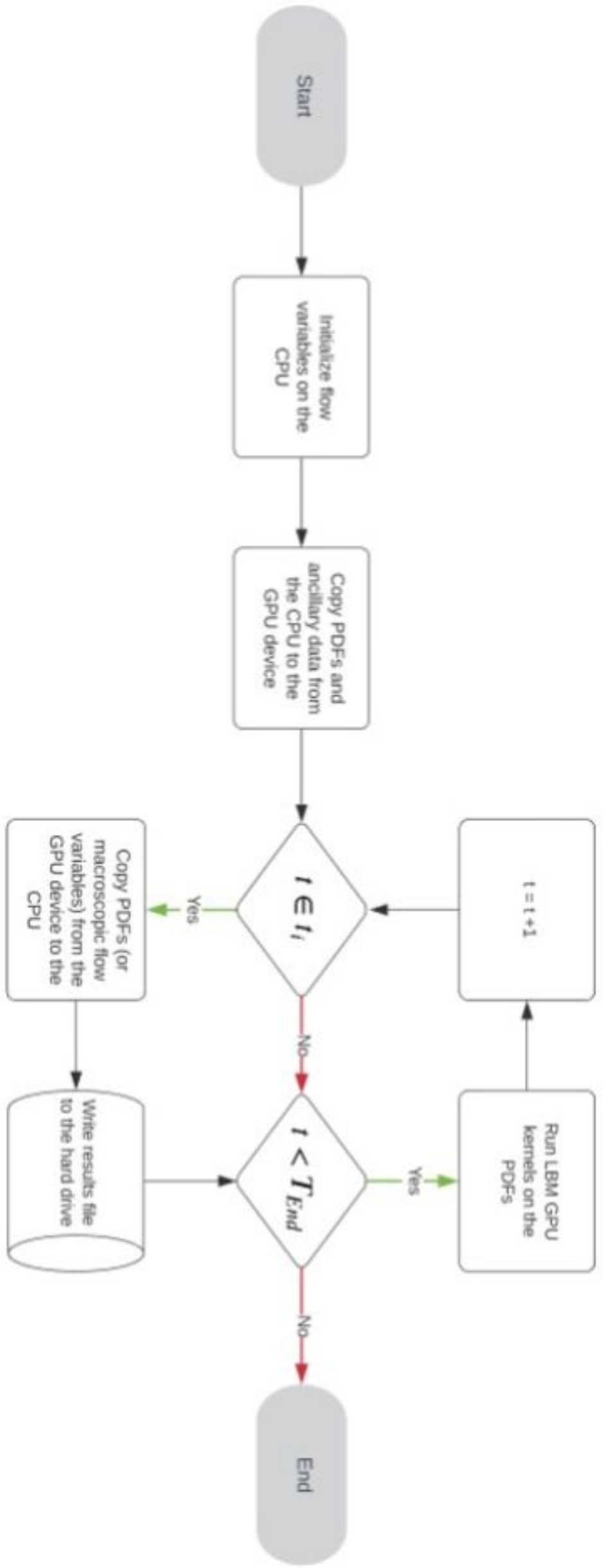
Start
Initialize flow variables on the CPU
Copy PDFs and ancillary data from the CPU to the GPU device
$t \in t_i$
Yes
No
Copy PDFs (or macroscopic flow variables) from the GPU device to the CPU
Write results file to the hard drive
$t < T_{End}$
Yes
No
Run LBM GPU kernels on the PDFs
t = t +1
End

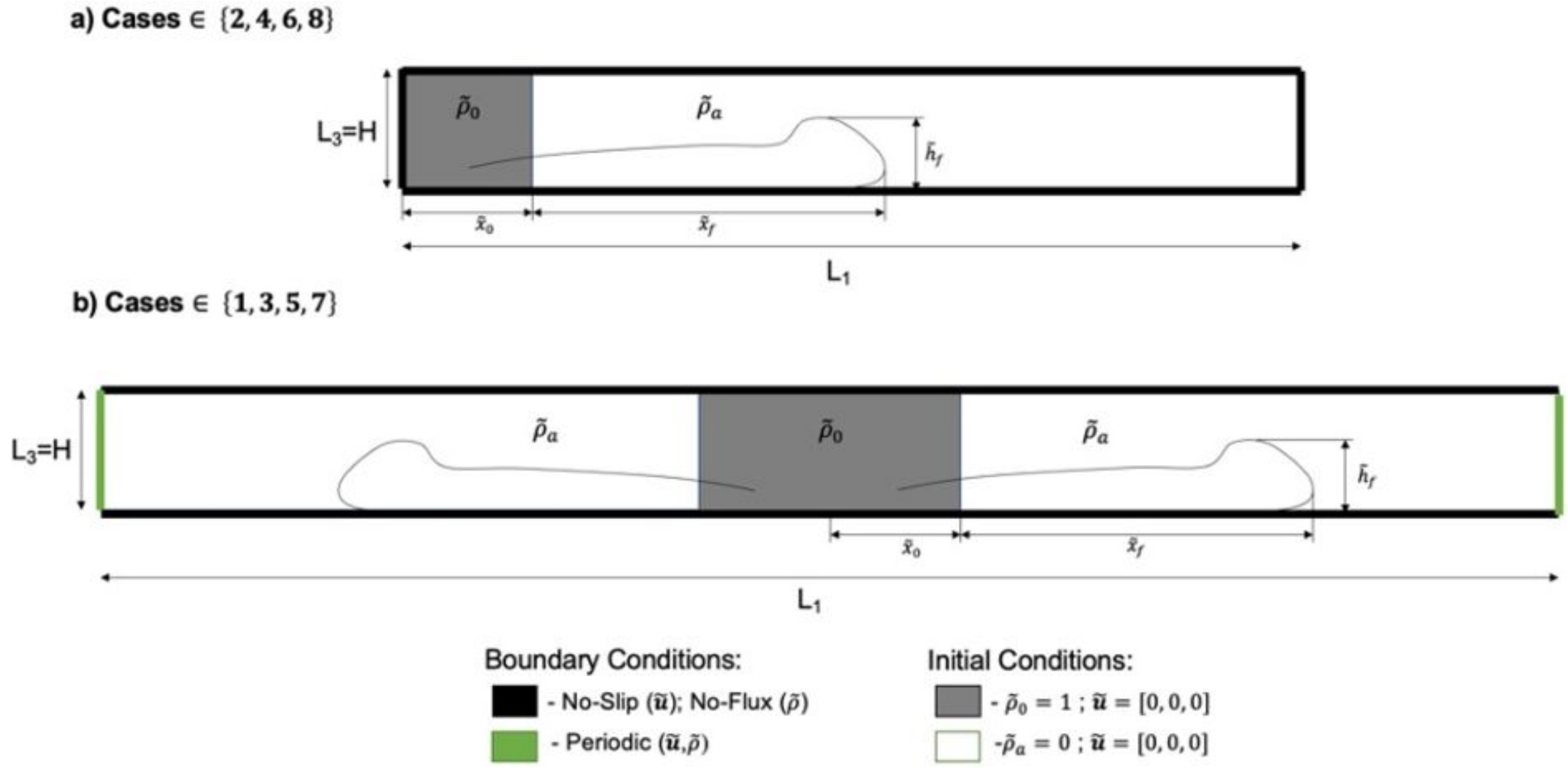
a) Cases ∈ {2, 4, 6, 8}
$L_3$=H
$\tilde{\rho}_0$
$\tilde{\rho}_a$
$\tilde{h}_f$
$\tilde{x}_0$
$\tilde{x}_f$
$L_1$
b) Cases ∈ {1, 3, 5, 7}
$L_3$=H
$\tilde{\rho}_a$
$\tilde{\rho}_0$
$\tilde{\rho}_a$
$\tilde{h}_f$
$\tilde{x}_0$
$\tilde{x}_f$
$L_1$
Boundary Conditions:
- No-Slip ($\tilde{\boldsymbol{u}}$); No-Flux ($\tilde{\rho}$)
- Periodic ($\tilde{\boldsymbol{u}}$,$\tilde{\rho}$)
Initial Conditions:
- $\tilde{\rho}_0 = 1$ ; $\tilde{\boldsymbol{u}} = [0,0,0]$
-$\tilde{\rho}_a = 0$ ; $\tilde{\boldsymbol{u}} = [0,0,0]$

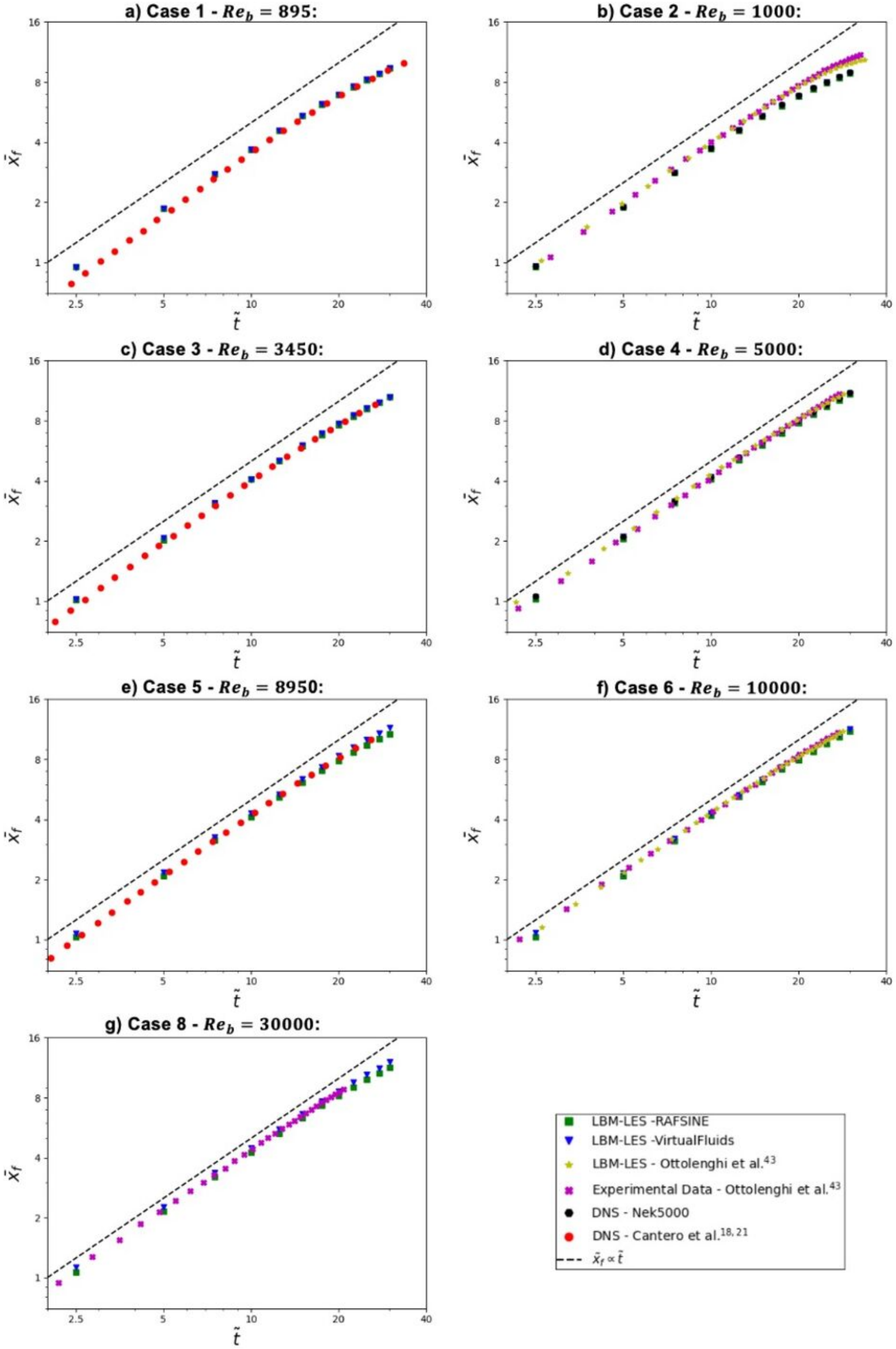
a) Case 1 - $Re_b = 895$:
b) Case 2 - $Re_b = 1000$:
c) Case 3 - $Re_b = 3450$:
d) Case 4 - $Re_b = 5000$:
e) Case 5 - $Re_b = 8950$:
f) Case 6 - $Re_b = 10000$:
g) Case 8 - $Re_b = 30000$:
$\tilde{x}_f$
$\tilde{t}$
LBM-LES -RAFSINE
LBM-LES -VirtualFluids
LBM-LES - Ottolenghi et al.[43]
Experimental Data - Ottolenghi et al.[43]
DNS - Nek5000
DNS - Cantero et al.[18,21]
$\tilde{x}_f \propto \tilde{t}$

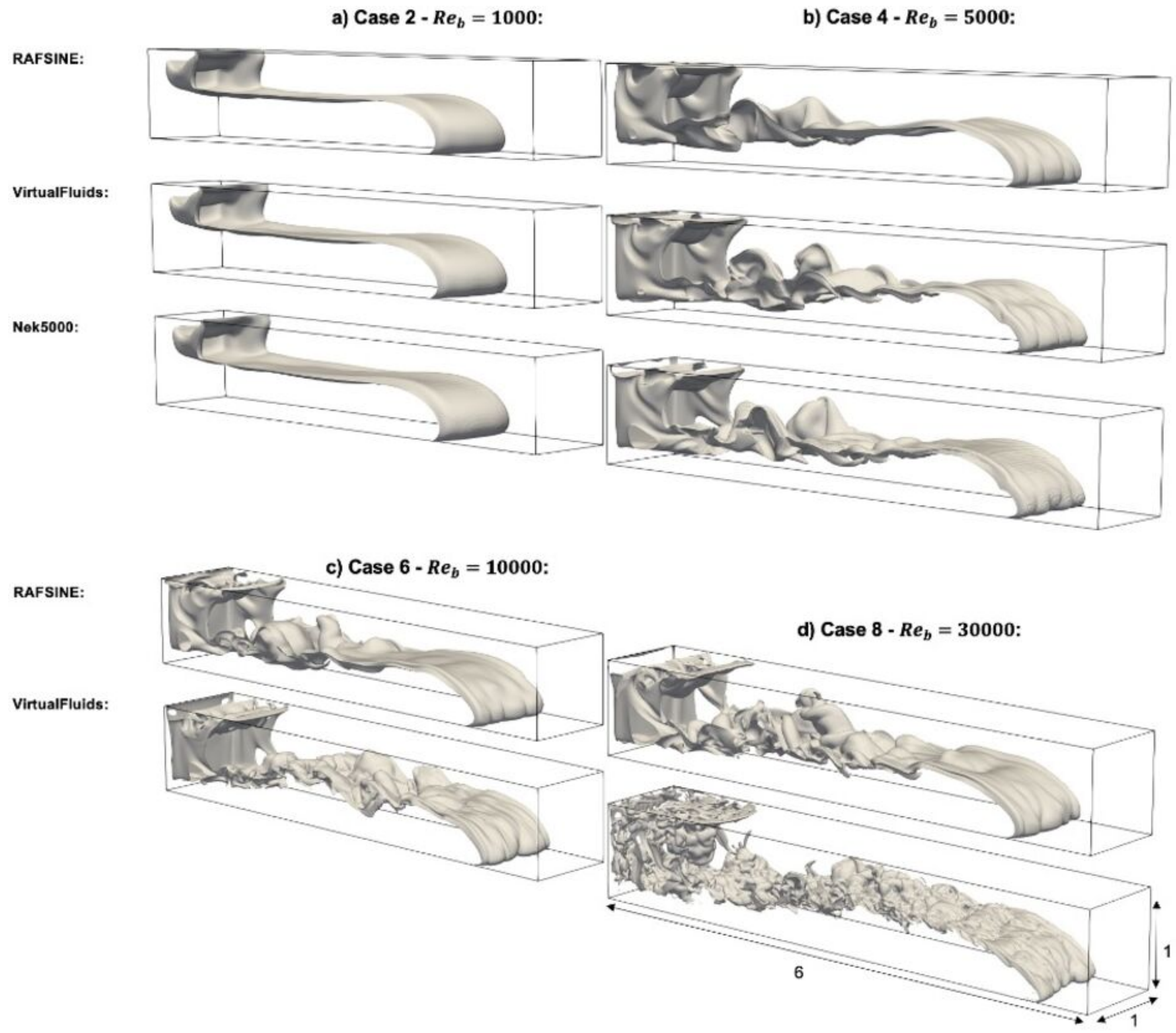
a) Case 2 - $Re_b = 1000$:
b) Case 4 - $Re_b = 5000$:
RAFSINE:
VirtualFluids:
Nek5000:
c) Case 6 - $Re_b = 10000$:
RAFSINE:
VirtualFluids:
d) Case 8 - $Re_b = 30000$:
6
1
1

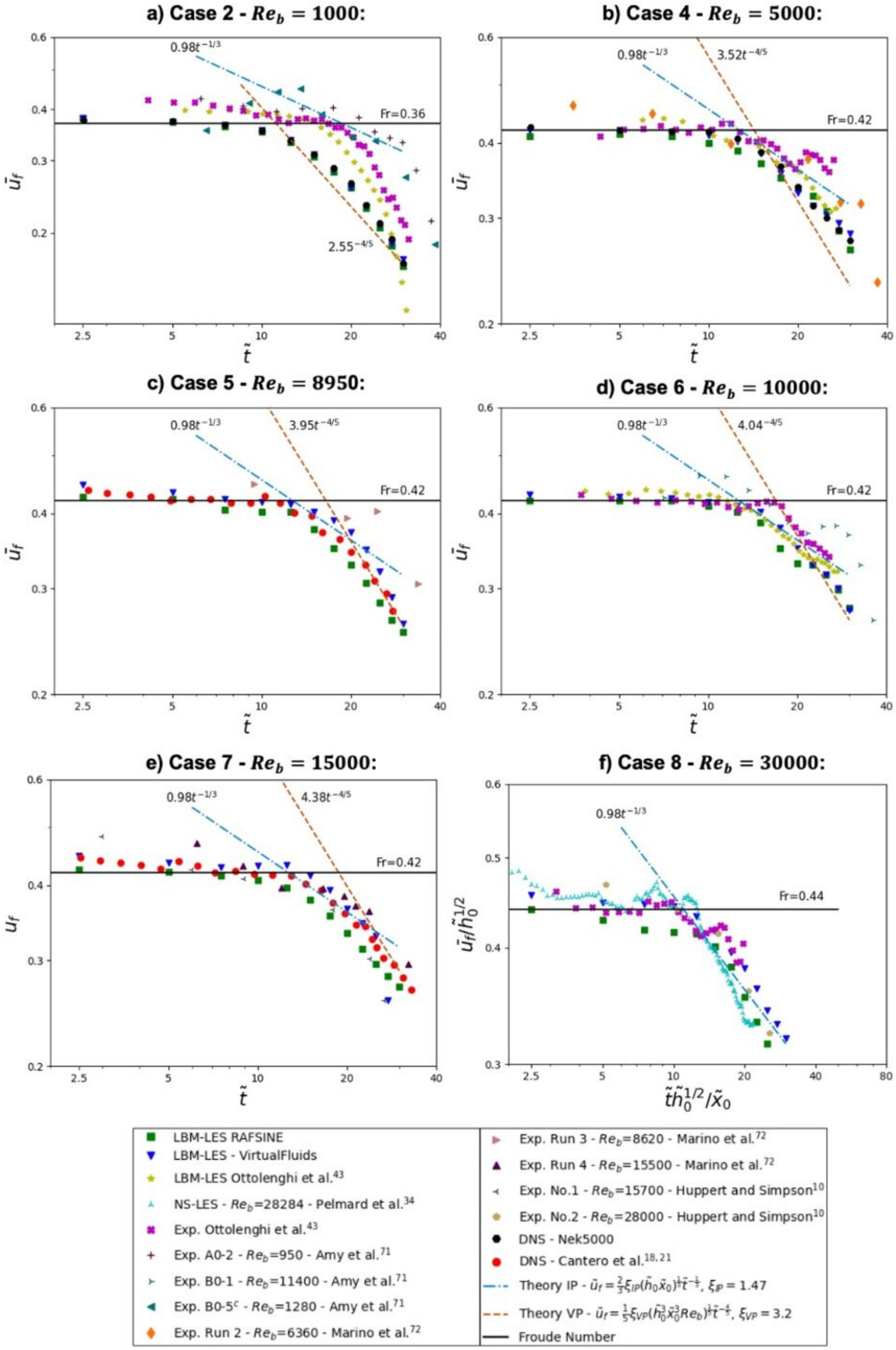
a) Case 2 - Re_b = 1000:
0.98t^-1/3
2.55^-4/5
Fr=0.36
b) Case 4 - Re_b = 5000:
0.98t^-1/3
3.52t^-4/5
Fr=0.42
c) Case 5 - Re_b = 8950:
0.98t^-1/3
3.95t^-4/5
Fr=0.42
d) Case 6 - Re_b = 10000:
0.98t^-1/3
4.04^-4/5
Fr=0.42
e) Case 7 - Re_b = 15000:
0.98t^-1/3
4.38t^-4/5
Fr=0.42
f) Case 8 - Re_b = 30000:
0.98t^-1/3
Fr=0.44
LBM-LES RAFSINE
LBM-LES - VirtualFluids
LBM-LES Ottolenghi et al.43
NS-LES - Re_b=28284 - Pelmard et al.34
Exp. Ottolenghi et al.43
Exp. A0-2 - Re_b=950 - Amy et al.71
Exp. B0-1 - Re_b=11400 - Amy et al.71
Exp. B0-5c - Re_b=1280 - Amy et al.71
Exp. Run 2 - Re_b=6360 - Marino et al.72
Exp. Run 3 - Re_b=8620 - Marino et al.72
Exp. Run 4 - Re_b=15500 - Marino et al.72
Exp. No.1 - Re_b=15700 - Huppert and Simpson10
Exp. No.2 - Re_b=28000 - Huppert and Simpson10
DNS - Nek5000
DNS - Cantero et al.18,21
Theory IP, ξ_IP = 1.47
Theory VP, ξ_VP = 3.2
Froude Number

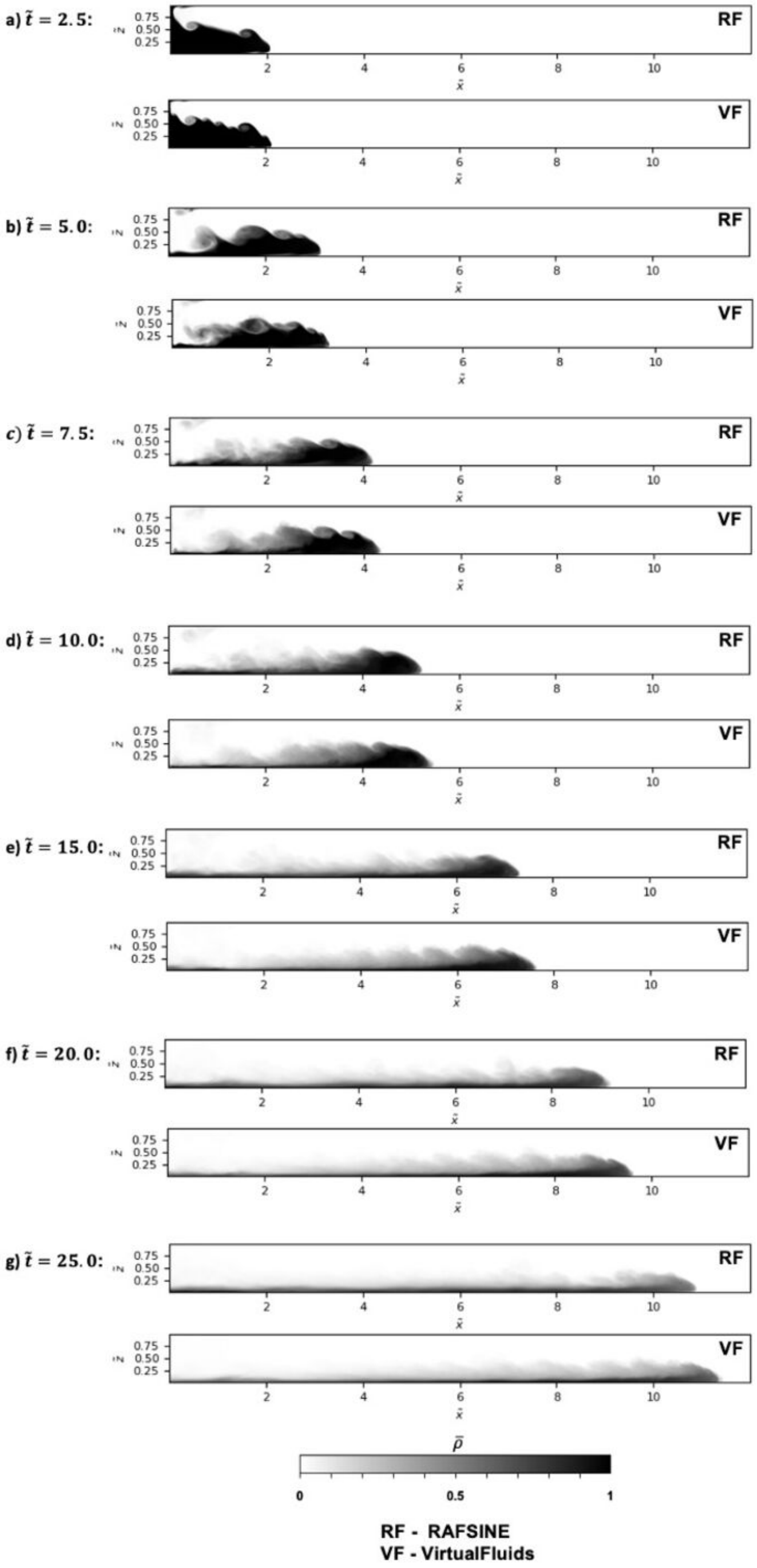
a) $\tilde{t} = 2.5$:
b) $\tilde{t} = 5.0$:
c) $\tilde{t} = 7.5$:
d) $\tilde{t} = 10.0$:
e) $\tilde{t} = 15.0$:
f) $\tilde{t} = 20.0$:
g) $\tilde{t} = 25.0$:
RF
VF
$\tilde{z}$
0.75
0.50
0.25
$\tilde{x}$
2
4
6
8
10
$\bar{\rho}$
0
0.5
1
RF - RAFSINE
VF - VirtualFluids

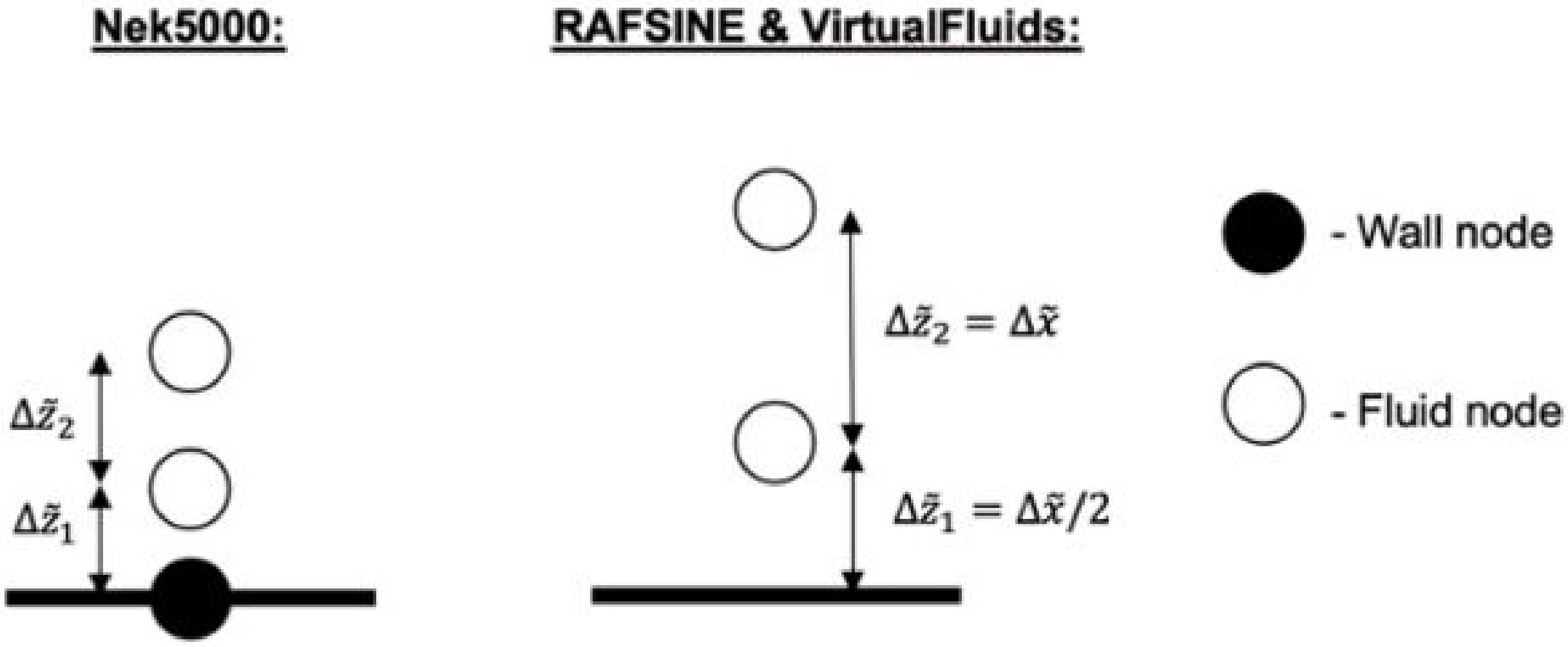

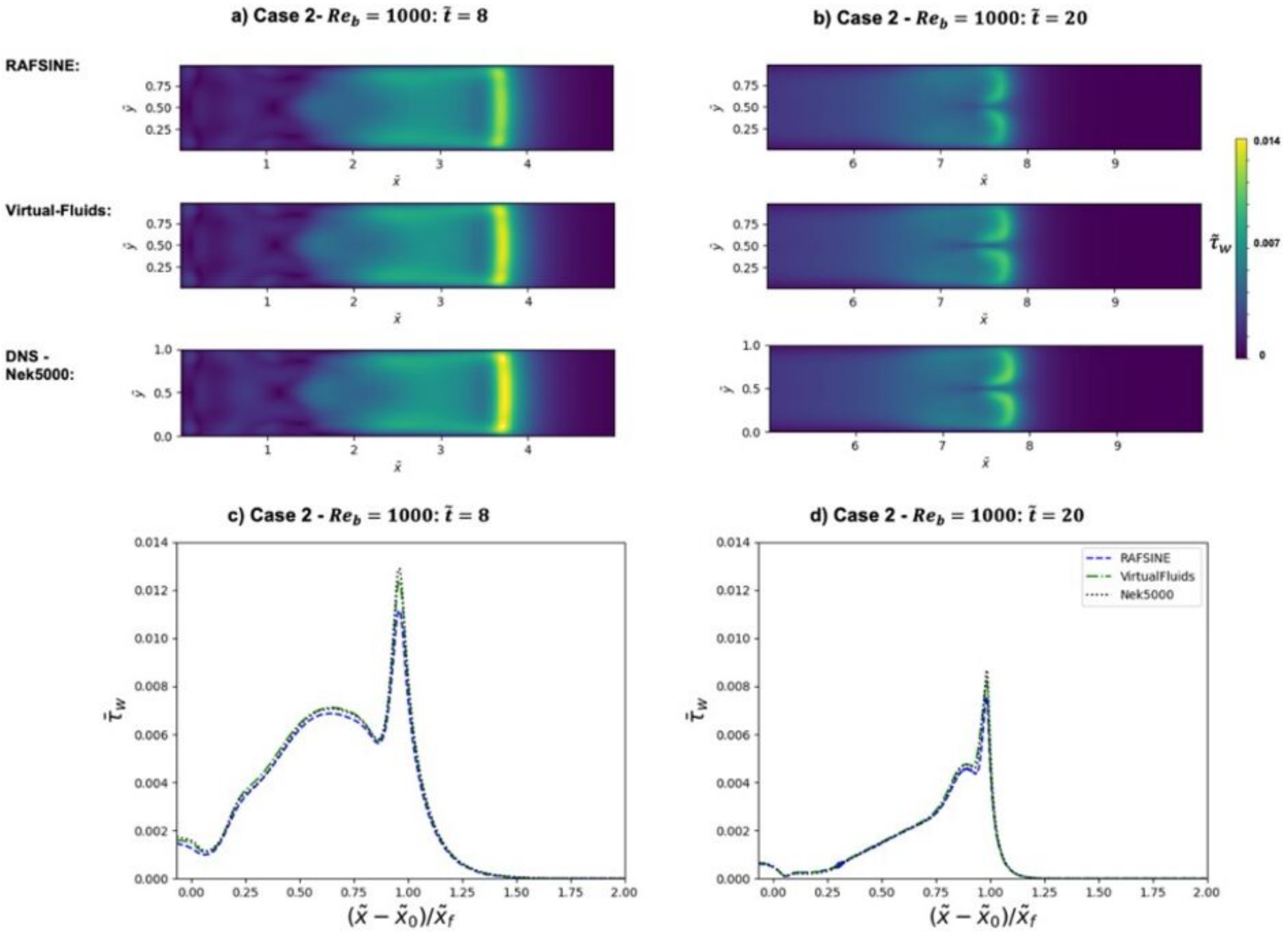
a) Case 2- $Re_b = 1000$: $\tilde{t} = 8$
b) Case 2 - $Re_b = 1000$: $\tilde{t} = 20$
RAFSINE:
Virtual-Fluids:
DNS -
Nek5000:
$\tilde{x}$
$\tilde{y}$
$\tilde{\tau}_w$
0.014
0.007
0
c) Case 2 - $Re_b = 1000$: $\tilde{t} = 8$
d) Case 2 - $Re_b = 1000$: $\tilde{t} = 20$
RAFSINE
VirtualFluids
Nek5000
$(\tilde{x} - \tilde{x}_0)/\tilde{x}_f$

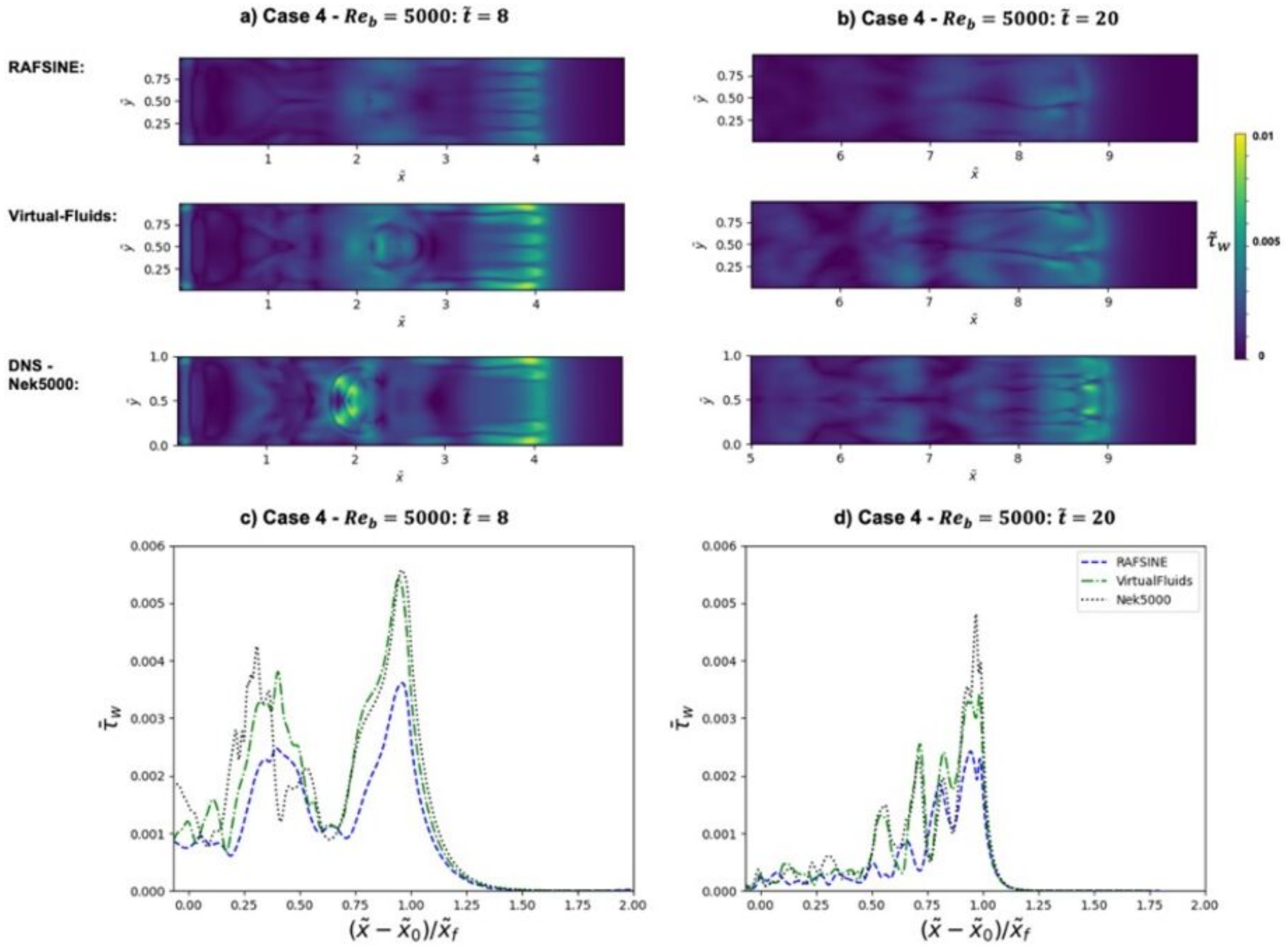
a) Case 4 - $Re_b = 5000$: $\tilde{t} = 8$
b) Case 4 - $Re_b = 5000$: $\tilde{t} = 20$
RAFSINE:
Virtual-Fluids:
DNS - Nek5000:
$\tilde{y}$
$\tilde{x}$
$\tilde{\tau}_w$
c) Case 4 - $Re_b = 5000$: $\tilde{t} = 8$
d) Case 4 - $Re_b = 5000$: $\tilde{t} = 20$
RAFSINE
VirtualFluids
Nek5000
$(\tilde{x} - \tilde{x}_0)/\tilde{x}_f$

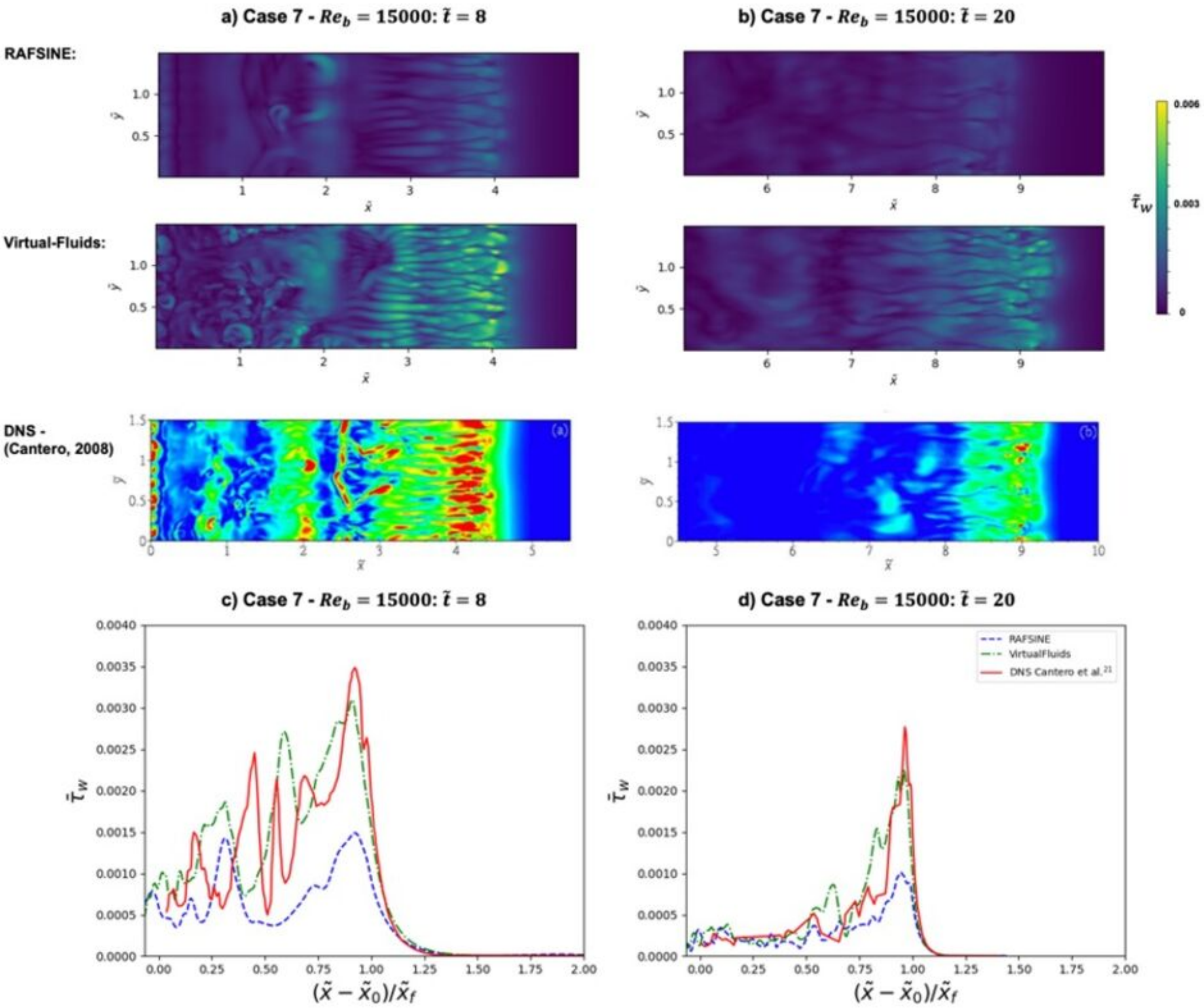

a) Case 7 - $Re_b = 15000$: $\tilde{t} = 8$
b) Case 7 - $Re_b = 15000$: $\tilde{t} = 20$
RAFSINE:
Virtual-Fluids:
DNS - (Cantero, 2008)
$\tilde{\tau}_w$
0.006
0.003
0
c) Case 7 - $Re_b = 15000$: $\tilde{t} = 8$
d) Case 7 - $Re_b = 15000$: $\tilde{t} = 20$
RAFSINE
VirtualFluids
DNS Cantero et al.[21]
$(\tilde{x} - \tilde{x}_0)/\tilde{x}_f$

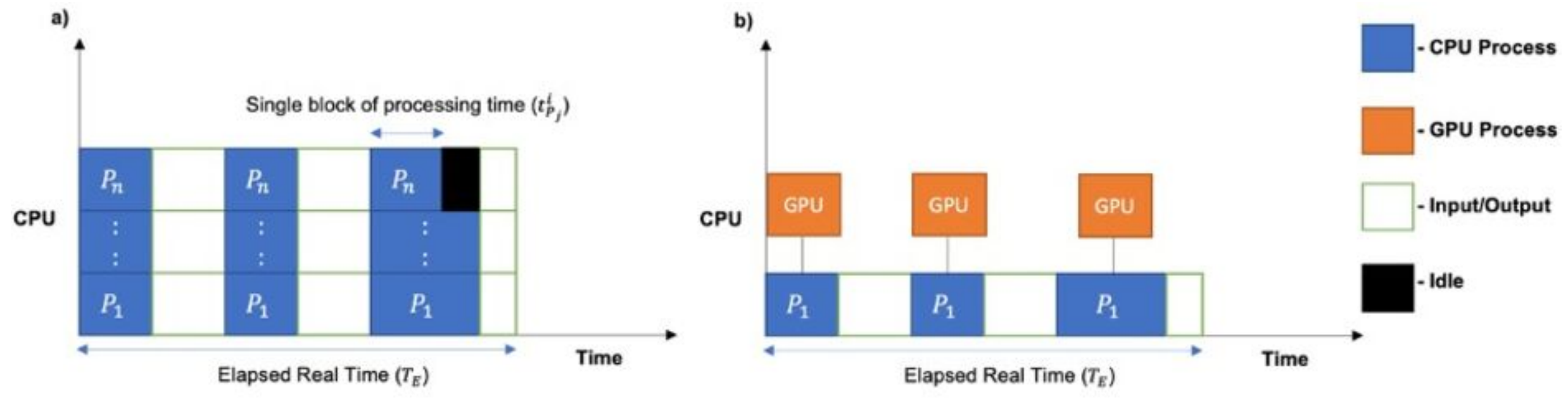

$$T_{CPU} = \sum_{i,j} t^i_{P_j} \qquad T_{CPU} > T_E$$

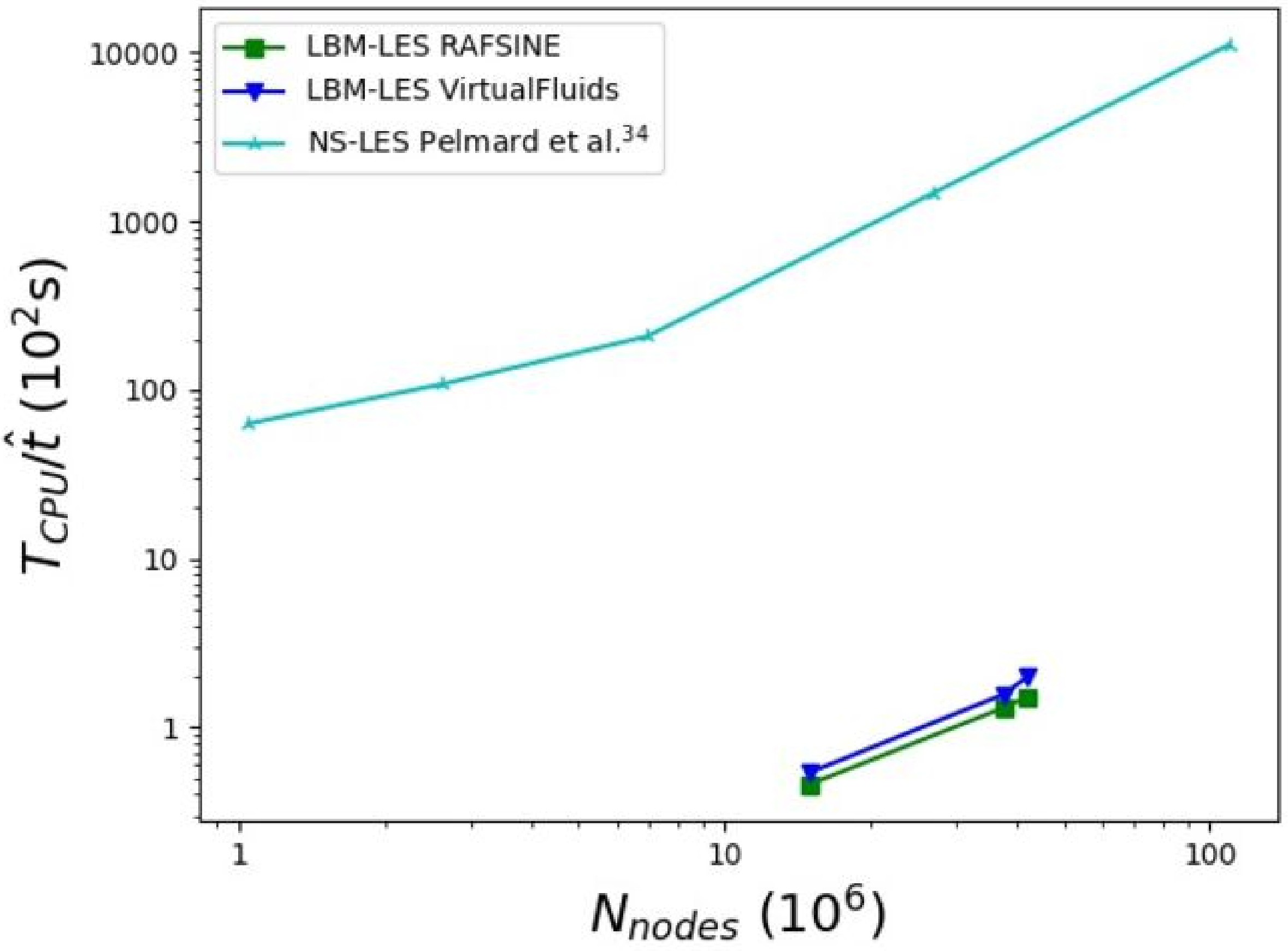